\PassOptionsToPackage{unicode}{hyperref}
\PassOptionsToPackage{hyphens}{url}
\PassOptionsToPackage{dvipsnames,svgnames,x11names}{xcolor}
\documentclass[
  11pt,
]{article}

\usepackage{amsmath,amssymb}
\usepackage{iftex}
\ifPDFTeX
  \usepackage[T1]{fontenc}
  \usepackage[utf8]{inputenc}
  \usepackage{textcomp} 
\else 
  \usepackage{unicode-math}
  \defaultfontfeatures{Scale=MatchLowercase}
  \defaultfontfeatures[\rmfamily]{Ligatures=TeX,Scale=1}
\fi
\usepackage{lmodern}
\ifPDFTeX\else  
\fi
\IfFileExists{upquote.sty}{\usepackage{upquote}}{}
\IfFileExists{microtype.sty}{
  \usepackage[]{microtype}
  \UseMicrotypeSet[protrusion]{basicmath} 
}{}
\makeatletter
\@ifundefined{KOMAClassName}{
  \IfFileExists{parskip.sty}{%
    \usepackage{parskip}
  }{
    \setlength{\parindent}{0pt}
    \setlength{\parskip}{6pt plus 2pt minus 1pt}}
}{
  \KOMAoptions{parskip=half}}
\makeatother
\usepackage{xcolor}
\usepackage[left=2.54cm,right=2.54cm]{geometry}
\ifx\paragraph\undefined\else
  \let\oldparagraph\paragraph
  \renewcommand{\paragraph}[1]{\oldparagraph{#1}\mbox{}}
\fi
\ifx\subparagraph\undefined\else
  \let\oldsubparagraph\subparagraph
  \renewcommand{\subparagraph}[1]{\oldsubparagraph{#1}\mbox{}}
\fi

\usepackage{longtable,booktabs,array}
\usepackage{calc} 
\usepackage{etoolbox}
\makeatletter
\patchcmd\longtable{\par}{\if@noskipsec\mbox{}\fi\par}{}{}
\makeatother
\IfFileExists{footnotehyper.sty}{\usepackage{footnotehyper}}{\usepackage{footnote}}
\makesavenoteenv{longtable}
\usepackage{graphicx}
\makeatletter
\def\maxwidth{\ifdim\Gin@nat@width>\linewidth\linewidth\else\Gin@nat@width\fi}
\def\maxheight{\ifdim\Gin@nat@height>\textheight\textheight\else\Gin@nat@height\fi}
\makeatother
\setkeys{Gin}{width=\maxwidth,height=\maxheight,keepaspectratio}
\makeatletter
\def\fps@figure{htbp}
\makeatother
\newlength{\cslhangindent}
\newlength{\csllabelwidth}
\newlength{\cslentryspacingunit} 
\newenvironment{CSLReferences}[2] 
 {
  \setlength{\parindent}{0pt}
  \ifodd #1
  \let\oldpar\par
  \def\par{\hangindent=\cslhangindent\oldpar}
  \fi
  \setlength{\parskip}{#2\cslentryspacingunit}
 }%
 {}
\usepackage{calc}

\usepackage[noblocks]{authblk}

\usepackage{lipsum} \usepackage{libertine}
\makeatletter
\@ifpackageloaded{caption}{}{\usepackage{caption}}
\AtBeginDocument{%
\ifdefined\contentsname
  \renewcommand*\contentsname{Table of contents}
\else
  \newcommand\contentsname{Table of contents}
\fi
\ifdefined\listfigurename
  \renewcommand*\listfigurename{List of Figures}
\else
  \newcommand\listfigurename{List of Figures}
\fi
\ifdefined\listtablename
  \renewcommand*\listtablename{List of Tables}
\else
  \newcommand\listtablename{List of Tables}
\fi
\ifdefined\figurename
  \renewcommand*\figurename{Figure}
\else
  \newcommand\figurename{Figure}
\fi
\ifdefined\tablename
  \renewcommand*\tablename{Table}
\else
  \newcommand\tablename{Table}
\fi
}
\@ifpackageloaded{float}{}{\usepackage{float}}
\floatstyle{ruled}
\@ifundefined{c@chapter}{\newfloat{codelisting}{h}{lop}}{\newfloat{codelisting}{h}{lop}[chapter]}
\floatname{codelisting}{Listing}

\makeatother
\makeatletter
\@ifpackageloaded{caption}{}{\usepackage{caption}}
\@ifpackageloaded{subcaption}{}{\usepackage{subcaption}}
\makeatother
\makeatletter
\@ifpackageloaded{tcolorbox}{}{\usepackage[skins,breakable]{tcolorbox}}
\makeatother
\makeatletter
\@ifundefined{shadecolor}{\definecolor{shadecolor}{rgb}{.97, .97, .97}}
\makeatother
\ifLuaTeX
  \usepackage{selnolig}  
\fi
\IfFileExists{bookmark.sty}{\usepackage{bookmark}}{\usepackage{hyperref}}
\IfFileExists{xurl.sty}{\usepackage{xurl}}{} 
\hypersetup{
  pdftitle={Reduced mobility? Urban exodus? Medium-term impacts of the COVID-19 pandemic on internal population movements in Latin American countries},
  pdfauthor={Francisco Rowe; Carmen Cabrera-Arnau; Miguel González-Leonardo; Andrea Nasuto; Ruth Neville},
  colorlinks=true,
  linkcolor={blue},
  filecolor={Maroon},
  citecolor={Blue},
  urlcolor={Blue},
  pdfcreator={LaTeX via pandoc}}

\title{\textbf{Reduced mobility? Urban exodus? Medium-term impacts of
the COVID-19 pandemic on internal population movements in Latin American
countries}}

\author[1]{Francisco Rowe}
\author[1]{Carmen Cabrera-Arnau}
\author[2, 3]{Miguel González-Leonardo}
\author[1]{Andrea Nasuto}
\author[1]{Ruth Neville}

\affil[1]{Geographic Data Science Lab, Department of Geography and
Planning, University of Liverpool, Liverpool, United Kingdom}
\affil[2]{Centre for Demographic Urban and Environmental Studies, El
Colegio de México, Ciudad de México, México}
\affil[3]{International Institute for Applied Systems Analysis,
Wittgenstein Centre, Vienna, Austria}

\date{}
\begin{document}
\maketitle
\begin{abstract}
The COVID‐19 pandemic has impacted the national systems of population
movement around the world. Existing work has focused on countries of the
Global North and restricted to the immediate effects of COVID-19 data
during 2020. Data have represented a major limitation to monitor change
in mobility patterns in countries in the Global South. Drawing on
aggregate anonymised mobile phone location data from Meta‐Facebook
users, we aim to analyse the extent and persistence of changes in the
levels (or intensity) and spatial patterns of internal population
movement across the rural-urban continuum in Argentina, Chile and Mexico
over a 26-month period from March 2020 to May 2022. We reveal an overall
systematic decline in the level of short- and long-distance movement
during the enactment of nonpharmaceutical interventions in 2020, with
the largest reductions occurred in the most dense areas. We also show
that these levels bounced back closer to pre-pandemic levels in 2022
following the relaxation of COVID-19 stringency measures. However, the
intensity of these movements has remained below pre-pandemic levels in
many areas in 2022. Additionally our findings lend some support to the
idea of an urban exodus. They reveal a continuing negative net balances
of short-distance movements in the most dense areas of capital cities in
Argentina and Mexico, reflecting a pattern of suburbanisation. Chile
displays limited changes in the net balance of short-distance movements
but reports a net loss of long-distance movements. These losses were,
however, temporary, moving to neutral and positive balances in 2021 and
2022.
\end{abstract}
\ifdefined\Shaded\renewenvironment{Shaded}{\begin{tcolorbox}[frame hidden, enhanced, breakable, boxrule=0pt, interior hidden, sharp corners, borderline west={3pt}{0pt}{shadecolor}]}{\end{tcolorbox}}\fi

\hypertarget{introduction}{%
\section{Introduction}\label{introduction}}

The COVID-19 pandemic triggered major changes in the patterns of human
population movement across the world. Nonpharmaceuthical interventions
to contain the spread of COVID-19, such as social distancing, lockdowns,
business and school closures contributed to sharp declines in internal
population movements within countries (Nouvellet et al. 2021). As the
virus spread globally in early 2020, COVID‐19 infection and mortality
rates were disproportionately higher in large and dense metropolitan
areas (Pomeroy and Chiney 2021), remote work reduced the need to commute
(Nathan and Overman 2020) and diminished retail activity decreased the
attractiveness and vibrancy of these areas (Florida, Rodr\'iguez-Pose, and
Storper 2021). In a gloomy prospect for urban centres, media headlines
suggested the death of cities gathering anecdotal evidence on an ``urban
exodus'', claiming sudden rises in the number of people moving from
large cities to suburbs and rural locales (Paybarah, Bloch, and Reinhard
2020; Marsh 2020).

Empirical evidence have examined these claims focusing on two key
dimensions of internal population movements i.e.~its level (or
intensity), and its spatial impacts on population redistribution.
Existing evidence reveals a sharp reduction in levels of overall
mobility, including short- and long-distance movements, and some support
for an urban exodus during the early days of COVID-2019 in 2020 as the
pandemic was unfolding and strict nonpharmaceuthical measures were
enacted (Rowe, González-Leonardo, and Champion 2023). Population losses
through internal population movements were documented for large core
metropolitan areas in Spain (González-Leonardo et al. 2022;
González-Leonardo and Rowe 2022), Germany (Stawarz et al. 2022), Sweden
(Vogiazides and Kawalerowicz 2022), Japan (Tønnessen 2021), Australia
(Perales and Bernard 2022), the United States (Ramani and Bloom 2021)
and Britain (Rowe et al. 2022). Suburban and outer parts of metropolitan
areas (as in the `donut effect' noted for US cities, see Ramani and
Bloom (2021)), as well as more distant rural areas, were identified as
gaining population.

While this research has contributed to advancing our understanding of
the impacts of COVID-19 on internal population movements in the Global
North, less is known about the extent and durability of population
movements to and from large cities during the pandemic in the Global
South and extending beyond 2020. Existing work has focused on countries
in the Global North and restricted to the immediate effects of COVID-19
data during 2020. Only two studies have analysed data extending to 2021
and they offer mixed evidence. For Spain, González-Leonardo and Rowe
(2022) report continuing population losses through internal migration in
large cities, such as Madrid and Barcelona. In Britain, however, Rowe et
al. (2022) record a recovery in the levels of internal movements in low
and high density areas converging to pre-pandemic levels.

A major limitation to capture national-scale changes in the patterns of
internal population movement across the rural-urban hierarchy in the
Global South during the pandemic has been the lack of suitable data.
Traditionally census and survey data are used to study human mobility
patterns at such scale. But, these data systems are not frequently
updated and suffer from slow releases, with census data for example
being collected over intervals of ten years in most Latin American
countries and with a considerable lag for the release of information
(CEPAL 2022). These systems thus lack the temporal granularity to
analyse population movements over short-time periods (Green, Pollock,
and Rowe 2021). Digital footprints (DFs) left as a result of social
interactions on digital platforms offer an unique source of information
to capture human population movement in the unfolding data revolution
(Rowe 2023). Particularly DF data from mobile phone applications have
become a prominent source to sense patterns of human mobility at higher
geographical and temporal resolution over in real time, and especially
in data-scarce environments in less developed countries.

Drawing on Meta-Facebook users' mobile phone location data, we aim to
examine the extent and persistence of changes in the levels (or
intensity) and spatial patterns of internal population movement across
the rural-urban continuum in Argentina, Chile and Mexico over a 26-month
period from March 2020 to May 2022. Specifically, we seek to address the
following set of questions:

\begin{itemize}
\item
  How have the levels of mobility changed across the rural-urban
  hierarchy over time?
\item
  To what degree have large cities experienced population loss through
  internal migration movements during the pandemic? What have been the
  main destinations for people leaving large cities?
\item
  To what extent have the intensity and potential negative balance of
  population movement from large cities have persisted over time?
\end{itemize}

We primarily focus on national capital cities. In Latin American
countries, these cities have dominated the national system of internal
population movement since the 1930s, although they have recorded
continuing balance of moderate population losses due to internal
migration over the last three decades (Bernard et al. 2017;
Rodr\'iguez-Vignoli and Rowe 2018). We could expect that the magnitude of
these population losses to have accelerated during the COVID-19
pandemic. We distinguish between short- and long-distance movement to
identify the potential sources of the patterns underpinning observed
changes in population movements during the pandemic.

Our work contributes to expanding existing knowledge by offering new
empirical evidence on the medium-term impacts of COVID‐19 on internal
population movement. First, it provides a systemic understanding of the
overall population movement system in Latin America. While prior work
has focused on daily mobility or internal migration, our study provides
a more complete representation analysing how short- and long-distance
mobility have evolved during COVID‐19. Second, we capture the evolution
of the internal population movements from 2020 to 2022. Data limitations
have prevented previous studies from extending their analysis to
understand the patterns of internal population movements beyond 2020.
Third, we contribute some of the first empirical evidence documenting
the patterns of internal population movement in a set of countries in
the Global South.

The rest of the paper is structured as follows. The next section
provides a brief review of recent research on internal population
movements during the COVID-19 pandemic, before providing an overview of
the main patterns of population movements within Latin American
countries. Section~\ref{sec-data} describes the data, and
Section~\ref{sec-methods} discusses the methods used in this study.
Section~\ref{sec-results} presents the key results from our analyses
before they are discussed in light of the existing literature in
Section~\ref{sec-discussion}, which also identifies key limitations and
potential avenues for future work. Section~\ref{sec-conclusion} provides
some concluding remarks.

\hypertarget{sec-background}{%
\section{Background}\label{sec-background}}

\hypertarget{the-impact-of-covid-19-on-internal-population-movements}{%
\subsection{The impact of COVID-19 on internal population
movements}\label{the-impact-of-covid-19-on-internal-population-movements}}

Globally, there is evidence that the COVID-19 pandemic constrained both
short- and long-distance movements within national boundaries (Nouvellet
et al. 2021; González-Leonardo, Rowe, and Fresolone-Caparrós 2022; Wang
et al. 2022; Rowe, González-Leonardo, and Champion 2023). Declines were
documented across the Global North during the first year of the pandemic
in the United States (Ramani and Bloom 2021), a number of European
countries, Japan and Australia (Rowe, González-Leonardo, and Champion
2023). The magnitude of these declines vary widely from 2.5\% in Spain
(González-Leonardo et al. 2022) to 8.5\% in Australia (Perales and
Bernard 2022). The shapest drops occurred when nonpharmaceutical
interventions, such as stay-at-home requirements, travel restrictions,
mobility restrictions, business and school closures were enacted.

These drops did not last very long. Existing evidence suggests that
volume of human mobility recovered pre-pandemic levels in most countries
following the ease of stringency measures, particularly lockdowns (Rowe,
Robinson, and Patias 2022). In Australia, declines in the intensity of
population movement were attributed to lockdowns, as well as increased
teleworking and a loss of labour market dynamism resulting from the
COVID-19-driven economic recession due to business closures and supply
shortages (Perales and Bernard 2022). In Latin America and the
Caribbean, evidence indicates that the implemention of social distancing
policies resulted in 10 percenge decrease in human mobility levels 15
days after they were enacted (Arom\'i et al. 2023). The highest declines
were recorded in Bolivia (19\%), Ecuador (17\%) and Argentina (16\%).
Smaller drops of 3\% were registered in Paraguay and Venezuela (Arom\'i et
al. 2023).

In Global North countries, evidence indicates that the COVID-19 pandemic
reshaped the spatial patterns of internal population movements.
Primarily the pandemic is believed to have trigger movements away from
large cities to sparsely populared areas, including suburban, coastal
and rural locations (Rowe, González-Leonardo, and Champion 2023).
Consistent findings have been reported from the United States (Ramani
and Bloom 2021), the United Kingdom (Rowe et al. 2022; Wang et al.
2022), Spain (González-Leonardo et al. 2022; González-Leonardo, Rowe,
and Fresolone-Caparrós 2022), Germany (Stawarz et al. 2022), Sweden
(Vogiazides and Kawalerowicz 2022), Norway (Tønnessen 2021), Australia
(Perales and Bernard 2022) and Japan (Fielding and Ishikawa 2021;
Kotsubo and Nakaya 2022). In the United States, Germany, Norway, Sweden
and Japan the predominant pattern was net migration losses in large
cities during 2020 but increases in their suburban areas. In the United
States, Ramani and Bloom (2021) labelled this phenomenon as ``donut
effect'', reflecting an increase net population loss through population
movement in metropolitan areas (urbanisation) and simultaneously a
rising net population gain in suburban rings (suburbanisation) and rural
areas (counterurbanisation).

In the United Kingdom, Spain and Japan, there is no evidence of a
``donut effect'' (Fielding and Ishikawa 2021; Rowe et al. 2022;
González-Leonardo, Rowe, and Vegas-Sánchez 2023). In these countries,
net population balances in suburbs did not show significant increases
during 2020 and 2021. Yet, population movements to large cities did
decline and movements to rural, sparsely populated areas increased,
leading to unusual large population gains in these areas. In Spain,
Sweden, Japan and Germany, attractive touristic, holiday and coastal
areas were popular destination for large city residents
(González-Leonardo, Rowe, and Fresolone-Caparrós 2022; Vogiazides and
Kawalerowicz 2022; Stawarz et al. 2022). These areas tend to be
locations with a prominent number of second and holiday homes,
suggesting that wealthy households who were able to work remotely left
large cities and settled in these areas (Haslag and Weagley 2021;
Tønnessen 2021).

The cumulative evidence has also suggested that while the patterns of
population movement were altered during the pandemic, the preexisting
predominant trends remained mostly unchanged. Most movements continued
to occur within and between urban areas, and the existing evidence
suggests that the observed sudden rises in movements from large cities
to rural location are not likely to endure the height of the pandemic
(Rowe, González-Leonardo, and Champion 2023). In the United Kingdom,
mobility levels had almost returned to prepandemic levels by August 2021
(Rowe et al. 2022; Wang et al. 2022). In Australia, changes in the
patterns of internal population movements due to COVID-19 had virtually
dissapered towards the end of 2020 (Perales and Bernard 2022). In Spain,
however, large cities such as Madrid and Barcelona have continued to
loss population through internal migration from 2020 to 2021
(González-Leonardo and Rowe 2022). While Barcelona had a record of net
migration losses prior to the pandemic, population losses due to
internal migration occurred during COVID-19. At the same time, unusually
high levels of counterurbanisation in Spain persisted over 2021
(González-Leonardo, Rowe, and Fresolone-Caparrós 2022). Thus, while the
levels and spatial patterns of internal population movements seem to
have resumed pandemic trends, key changes seem to have endure the height
of the COVID-19 pandemic. The long-term effects of COVID-19 remain to be
established.

Overall, existing work has contributed to better understand how internal
population movements across the rural-urban hierarchy were affected by
the COVID-19 pandemic in the Global North. However, less is known about
COVID-19 impacts on movements between cities, suburbs and rural areas in
the Global South and the durability of resulting changes. Yet, that is
not to say that evidence does not exist. Evidence based on small surveys
carried out in India (Irudaya Rajan, Sivakumar, and Srinivasan 2020) and
South Africa (Ginsburg et al. 2022) in 2020 suggests that flows from
large cities to less populated areas increased as a result of the return
of workers to their hometown as business closures and lockdowns were
enacted, At the same time, movements to cities diminished (Irudaya
Rajan, Sivakumar, and Srinivasan 2020). Both studies suggest that the
economic downturn caused by nonpharmaceutical interventions during the
pandemic (Ghosh, Seth, and Tiwary 2020) resulted in a reduction in
inflows of workers to cities, as well as an increase in the number of
return moves due to a rise in unemployment. This evidence suggests that
vulnerable populations in the Global South seem to have played a key
role in shaping the movements to and away from large cities during the
pandemic.

Yet, this evidence differ from recent work which points to an increase
in population movements made by wealthy individuals from large cities to
less densely populated areas in Global South countries during the first
wave of COVID-19. Lucchini et al. (2023) provide evidence analysing data
from Brazil, Colombia, Indonesia, Mexico, Philippines and South Africa.
They estimated that residents from high-wealthy neighborhoods were 1.5
times more likely to leave cities compared to those from low-wealthy
areas. This pattern is consistent with evidence from Chile indicating an
over-representation of high-income individuals in movements out of the
Greater Santiago Area during the pandemic (Elejalde et al. 2023).
Wealthy individuals are believed to have the resources to move, be able
to work remotely, and own second residences in popular holiday
destinations. Low-income households, on the other hand, may be less
movable as they may lack the financial resources to move. Also, a large
proportion of low-income families work in service or informal sectors of
the economy and jobs in these sectors require in-person, face-to-face
interactions.

Overall, there is little evidence analysing the patterns of internal
population movement in Global South countries. Though, evidence suggests
distinctive patterns in the selectivity of movements occurring in Latin
American countries. Yet, we are still to better understand the magnitude
and durability of changes to internal population movements across the
urban-rural hierarchy. Here, we use mobile phone data to analyse the
effect of COVID-19 on the patterns of internal population movements in
Argentina, Chile and Mexico. Before, we review literature on internal
population movements in Latin America, to offer some broader context
about the predominant patterns of mobility that were at work before the
unfolding of the COVID-19 pandemic.

\hypertarget{preexisting-patterns-of-population-movement}{%
\subsection{Preexisting patterns of population
movement}\label{preexisting-patterns-of-population-movement}}

Currently, Latin America has the highest urbanisation rate in the world
after North America, totaling 81\% (Nations" 2019). In Latin American
countries, population is highly concentrated in a few urban centres,
particularly in large cities with more than one million inhabitants,
where half of urban residents (Pinto da Cunha 2002; A. E. Lattes,
Rodr\'iguez, and Villa 2017). High urbanisation rates have been the result
of high levels of population redistribution from rural settlements to
cities, mostly during a fast phase of industrialisation between the
early 1950s to late 1970s (Firebaugh 1979; A. Lattes 1995). During these
decades, considerable population gains were mainly recorded in capital
cities (J. Sobrino 2012).

Since the 1980s, however, movements to metropolitan areas in Latin
America have diminished (Chávez Galindo et al. 2016). This has been a
natural consequence as rural population stocks depleted (Chávez Galindo
et al. 2016), coupled with a transition from an industrial substitution
economic strategy to free-market model led to a trend of population
deconcentration away from large cities, such as Santiago in Chile
(González Ollino and Rodr\'iguez Vignoli 2006; Rowe, 2013.) or Mexico City
(Jaime Sobrino 2006). This transition has resulted in a reduction in the
volume of long-distance movements to capital cities, leading to
population losses in some cities, like Santigo (Rowe, 2013.).

As a result, middle-sized cities became more attractive population
centres. They benefitted from increasing domestic and foreign investment
in export-oriented industries and tourism activities, leading to
geographic economic dispersal and subsequent migration inflows (Brea
2003; Pérez-Campuzano 2013; Chávez Galindo et al. 2016). Over the two
past decades, movements between cities have dominated the internal
migratory system in Latin American countries (Bernard et al. 2017;
Rodr\'iguez-Vignoli and Rowe 2018; Nations" 2019). About 80\% of internal
migrants moved between cities, according to the 2010-11 census round
(Rodr\'iguez Vignoli 2017). Medium-sized cities from 500.000 to 1 million
residents record the largest population gains due to internal migration
(Rodr\'i\'iguez Vignoli 2017). Large cities with more than 1 million
residents registered balanced rates, while small cities with less than
500.000 inhabitants recorded population losses through internal
migration (Rodr\'iguez-Vignoli and Rowe 2018).

Latin American cities have shown significant growth in terms of land
development in their urban peripheries. Since the 1970s, large cities,
such as Santiago de Chile, Buenos Aires and Mexico City have experienced
rapid suburbanisation (Graizbord and Acuña 2007; Chávez Galindo et al.
2016). Suburbanisation flows comprise middle- and high-class families
moving from cities to highly segregated areas on the periphery (Borsdorf
2003; Rodr\'iguez Vignoli 2019). Though, some peripheral areas have also
served home for low-income households settling in informal housing
(Janoschka 2002; Rodr\'iguez Vignoli and Rowe 2017). Peripheral areas have
thus become major residential areas for metropolitan residents serving
as satellite neighbourhoods for daily commuters to core areas of capital
cities (Chávez Galindo et al. 2016). Next we describe the data we used
to examine the extent to which the existing patterns of population
movement have been altered during the COVID-19 pandemic.

\hypertarget{sec-data}{%
\section{Data}\label{sec-data}}

\hypertarget{meta-facebook-data}{%
\subsection{Meta-Facebook Data}\label{meta-facebook-data}}

To capture population movements during the COVID-19 pandemic, we used
anonymised aggregate mobile phone location data from Meta users for
Argentina, Chile and Mexico, covering an 26-month period from March 2020
to May 2022. We used two datasets Facebook Movements and Facebook
Population created by Meta and accessed through their Data for Good
Initiative
(\href{https://dataforgood.facebook.com/}{https://dataforgood.facebook.com}).
The data are built from Facebook app users who have the location
services setting turned on on their smartphone. Prior to releasing the
datasets, Meta ensures privacy and anonymity by removing personal
information and applying privacy-preserving techniques (Maas et al.
2019). Small-count dropping is one of these techniques. A data entry is
removed if the population or movement count for an area is lower than
10. The removal of small counts may mean that population counts in small
sparsely populated areas are not captured. A second technique consists
in adding a small undisclosed amount of random noise to ensure that it
is not possible to ascertain precise, true counts for sparsely populated
locations. Third, spatial smoothing using inverse distance-weighted
averaging is also applied is applied to produce a smooth population
count surface.

The Facebook Movements dataset offers information on the total number of
Facebook users moving between and within spatial units in the form of
origin-destination matrices. The Facebook Population dataset offers
information on the number of active Facebook users in a spatial unit at
a given point in time. Both datasets are temporally aggregated into
three daily 8-hour time windows (i.e.~00:00-08:00, 08:00-16:00 and
16:00-00:00). The datasets include a baseline capturing the number of
Facebook user population or movement before COVID-19 based on a 45-day
period ending on March 10th 2020. The baseline is computed using an
average for the same time of the day and day of the week in the period
preceding March 10th. For instance, the baseline for Monday 00:00-08:00
time window is obtained by averaging over data collected on Mondays from
00:00 to 8:00 for a 45 day period. Details about the baseline can be
found in Maas et al. (2019).

The Bing Maps Tile System developed by Microsoft (Microsoft) is used a
spatial framework to organise the data. The Tile System is a geospatial
indexing system that partitions the world into tile cells in a
hierarchical way, comprising 23 different levels of detail (Microsoft).
At the lowest level of detail (Level 1), the world is divided into four
tiles with a coarse spatial resolution. At each successive level, the
resolution increases by a factor of two. The data that we used are
spatially aggregated into Bing tile levels 13. That is about 4.9 x 4.9km
at the Equator (Maas et al. 2019).

\hypertarget{covid-19-stringency-data}{%
\subsection{COVID-19 stringency data}\label{covid-19-stringency-data}}

We used the stringency index as a measure of the level of
nonpharmaceutical interventions to COVID-19, such as social distancing
and lockdowns. We used these data to understand how the level and
patterns of population movement changed in response to the
implementation of various nonpharmaceutical interventions during
COVID-19. The stringency index ranges from 0 to 100, with 100 being the
value corresponding to the most strict scenario. The stringency index
were retrieved from the COVID-19 government response tracker
{[}https://www.bsg.ox.ac.uk/research/research-projects/covid-19-government-response-tracker{]}(https://www.bsg.ox.ac.uk/research/research-projects/covid-19-government-response-tracker).
See Hale et al. (2021), for more information.

\hypertarget{worldpop-population-data}{%
\subsection{WorldPop population data}\label{worldpop-population-data}}

We also used data from WorldPop (Tatem 2017) to measure the degree of
population density of local areas. WorldPop offers open access gridded
population estimates at a resolution of 3 and 30 arc-seconds
approximately 100m and 1km at the Equator, respectively. WoldPop
produces these gridded datasets using a top-down (i.e.~dissagregating
administrative area counts into smaller grid cells) or bottom-up
(i.e.~interpolating data from counts from sample locations into grid
cells) approaches. The data are in raster and tabular data formats. We
gridded population data at 1km resolution in raster format. Data at this
level of resolution are needed to spatially integrate our Facebook
population and movement data.

\hypertarget{sec-methods}{%
\section{Methods}\label{sec-methods}}

\hypertarget{sec-methods1}{%
\subsection{Population density classification}\label{sec-methods1}}

We sought to analyse the extent and durability of changes in internal
population movements across the rural-urban continuum. We thus moved
away from a rural-urban binary framework to understand population
movements and capture the diversity of contemporary human settlement
systems. Champion and Graeme (2004) and Rowe et al. (2019) serve as
excellent references summarising these arguments. We focused on
understanding the patterns of population movement across the rural-urban
continuum, particularly population exchanges between dense urban cores,
suburbs, medium-size areas and rural locations. Following Rowe et al.
(2019), we measured population density to capture the rural-urban
continuum.

For individual countries, we spatially aggregated these the WorldPop
data into the Bing tiles to correspond with our Facebook population data
using spatial intersection operators and area weighted centroids. Area
weighted centroids are particularly useful to allocate grids on coastal
regions. We then classified the WorldPop population into population
density deciles, identifiying population density classes, with 1
capturing the least dense areas to 10 representing the most dense areas.
To this end, we employed the Jenks natural breaks classification method
(Jenks 1967).

\begin{figure}

{\centering \includegraphics{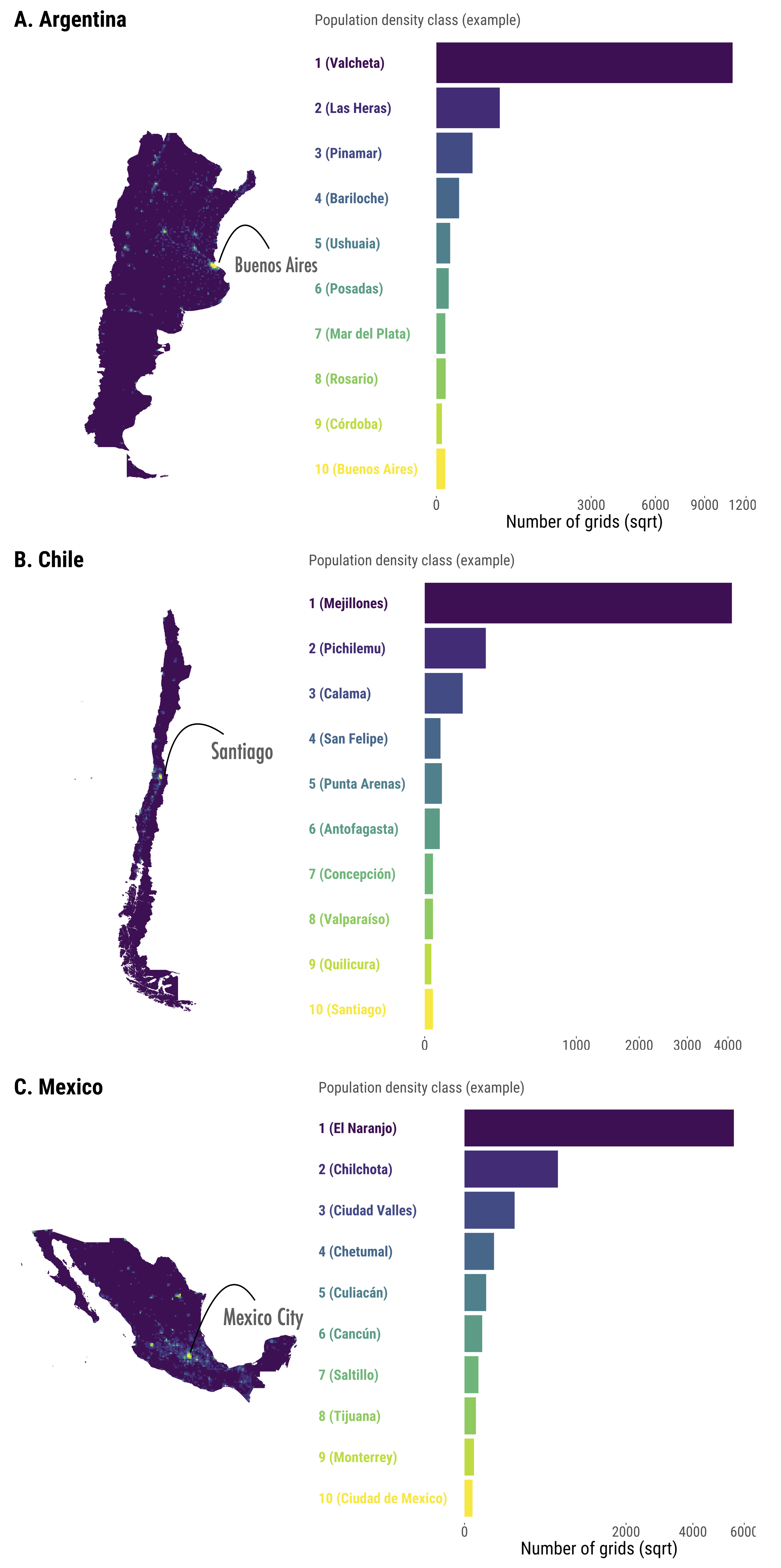}

}

\caption{\label{fig-map}Population density classes at Microsoft Bing
tile level and number of grids in each population density class.}

\end{figure}

Figure~\ref{fig-map} displays a map with our classification of Bing
tiles coloured according to the population density classes, and a bar
plot reporting the number of Bing tiles in each population classes. The
bar plot also shows example areas in bracket of the areas included in
each category, to provide an idea of the type of places that they
include. A key observation to recall for the remainder of the paper is
the fact that the most dense class 10 exclusively includes the core
metropolitan areas of the national capitals (i.e.~Buenos Aires in
Argentina, Santiago in Chile, and Mexico City in Mexico). Class 9 areas
tend to represent highly dense, suburban metropolitan areas around these
national capitals, and class 1 areas tend to be small, sparely populated
areas on coastal regions or around natural parks, natural reserves,
agricultural and forest areas. Also note that the population density
ranges for each density class differ by country. The logic of this
classification is that we expect tiles belonging to the same category in
individual countries have a similar function in the national settlement
system.

\hypertarget{tile-based-mobility-metrics}{%
\subsection{Tile-based mobility
metrics}\label{tile-based-mobility-metrics}}

\emph{Intensity of movement}

We sought to measure two key dimensions of population movements: its
level or intensity, and spatial impact (M. Bell et al. 2002). We
measured changes in the intensity of movement by quantifying the
variation in the number of population movements by population density
category in relation to the baseline period. We focused on two
distinctive months: May 2020 and March 2022 representing a period during
the height of the COVID-19 pandemic when a range of stringency measures
were in place, and a period at a very advanced stage of the pandemic
after mass vaccination has been rolled out and stringency measures had
been removed. Specifically, we used the percentage change in the
intensity of a population movement in reference to the reference period,
as provided in the Facebook Movement datasets (Maas et al. 2019):

\begin{equation}\protect\hypertarget{eq-eq1}{}{
Change (\%) = \Biggl( \frac{n_{crisis}}{n_{baseline}}-1 \Biggl) \times 100
}\label{eq-eq1}\end{equation}

where \(n_{crisis}\) corresponds to the number of people moving; and,
\(n_{baseline}\) corresponds to the number of people that would be
expected to move in the baseline pre-pandemic period at the same time of
the day, on the same day of the week, and between the same origin and
destination tiles. A positive score indicates an increase in the extent
of population movement relative to baseline levels. A negative score
represents a decrease, while a zero score denotes no change.

We grouped the resulting percentage change metrics by population density
class for places of origin, and generate boxplots to capture the
distribution of changes across these classes. We also considered the
relationship between these changes in mobility and degree of stringency
during the pandemic.

\emph{Spatial impact of movement}

We also sought to examine the degree of population redistribution
through mobility. Particularly we aimed to explore the extent to which
people have moved from capital cities down in the rural-urban hierarchy.
To this end, we computed the net balance between inflows and outflows:

\begin{equation}\protect\hypertarget{eq-eq2}{}{
netflows = inflows - outflows
}\label{eq-eq2}\end{equation}

where: \(inflows\) and \(outflows\) represent the total number of
movement in and out of an area. We computed and aggregated the net
balance by each tile to produce an overall metric for individual
population density class. Thus, the net balance for population density
class 10 areas represent the overall net balance in capital cities.
Also, because we computed the net balance for individual areas at three
windows during a day, we avoid focusing on daily mobility. Over the
course of a day, positive and negative net balances of daily mobility
would cancel out, and hence capture moves involving one-night stay.

\hypertarget{local-and-long-distance-movement}{%
\subsection{Local and long-distance
movement}\label{local-and-long-distance-movement}}

To further mitigate the effect of daily and short-distance moves, we
distinguish between short- and long-distance movements. In the
literature, short-distance moves are often used to capture residential
mobility, while long-distance moves are used to represent internal
migration (Niedomysl 2011). Short-distance moves are associated
primarily with housing reasons, while long-distance moves are driven
primarily by labour market reasons. For the metrics described above, we
therefore considered only flows with a Euclidean distance between
origins and destinations greater than 0km. We defined short-distance
moves as trips of less 100km, and long-distance moves as trips covering
100km or more.

In terms of expectations, we expected to observe an overall decrease in
mobility as a consequence of stringency measures especially during the
early stages of the COVID-19 pandemic. During this period, we also
expected to find a decrease in movement in and to metropolitan cores and
an increase in mobility to suburban areas or satellite cities,
reflecting changes in commuting patterns. We expected to observe
negative net balances of movements in big cities and positive balances
down the rural-urban hierarchy, particularly in less dense, coastal
towns, rural and lake areas.

\hypertarget{sec-results}{%
\section{Results}\label{sec-results}}

\hypertarget{changes-in-mobility-in-may-2020-and-in-march-2022}{%
\subsection{Changes in mobility in May 2020 and in March
2022}\label{changes-in-mobility-in-may-2020-and-in-march-2022}}

\begin{figure}

{\centering \includegraphics{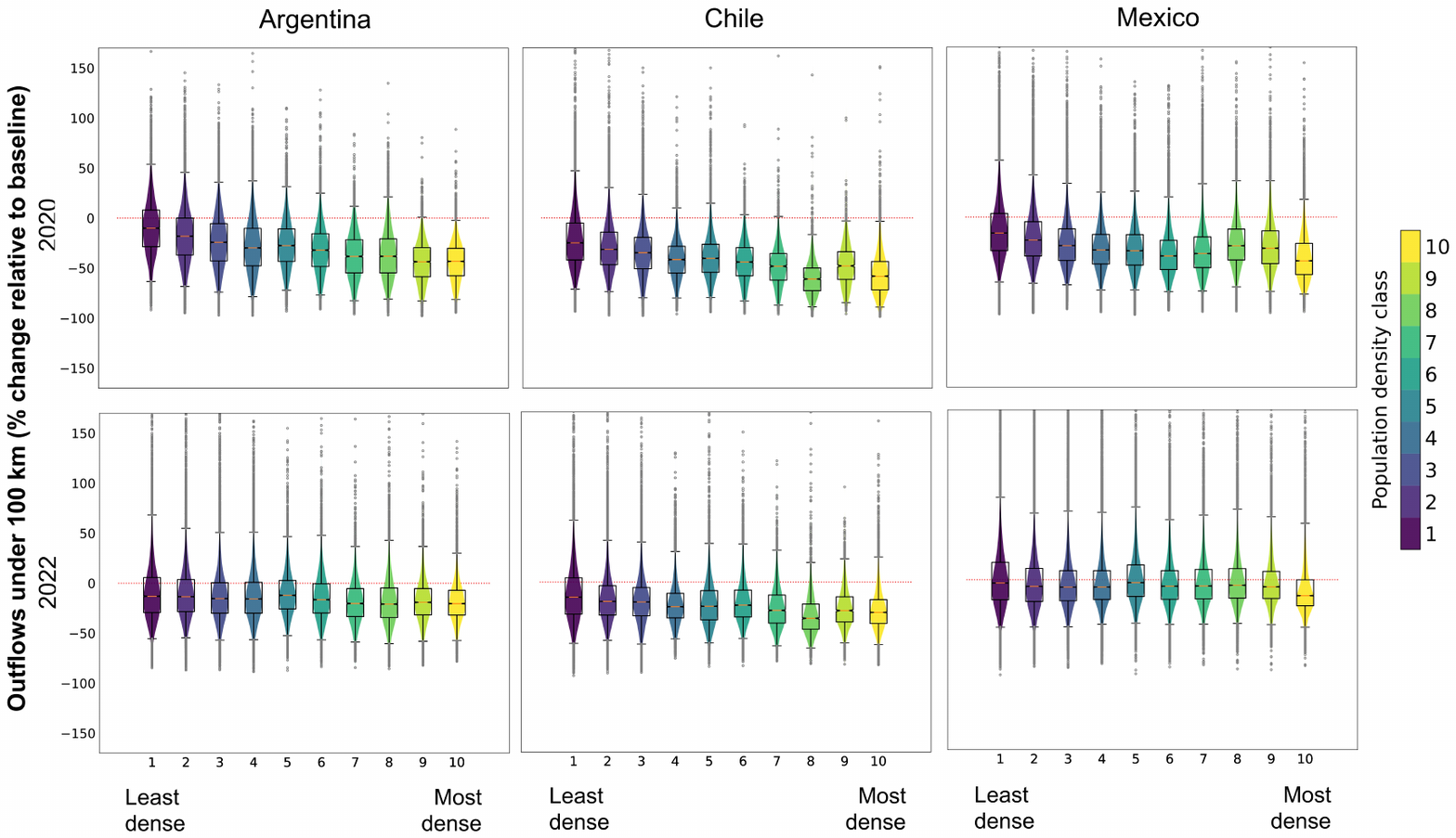}

}

\caption{\label{fig-outflowsU100}Changes in mobility flows during May
2020 and March 2022 by population density class, relative to baseline
period. Short-distance (\textless100km) movements.}

\end{figure}

\begin{figure}

{\centering \includegraphics{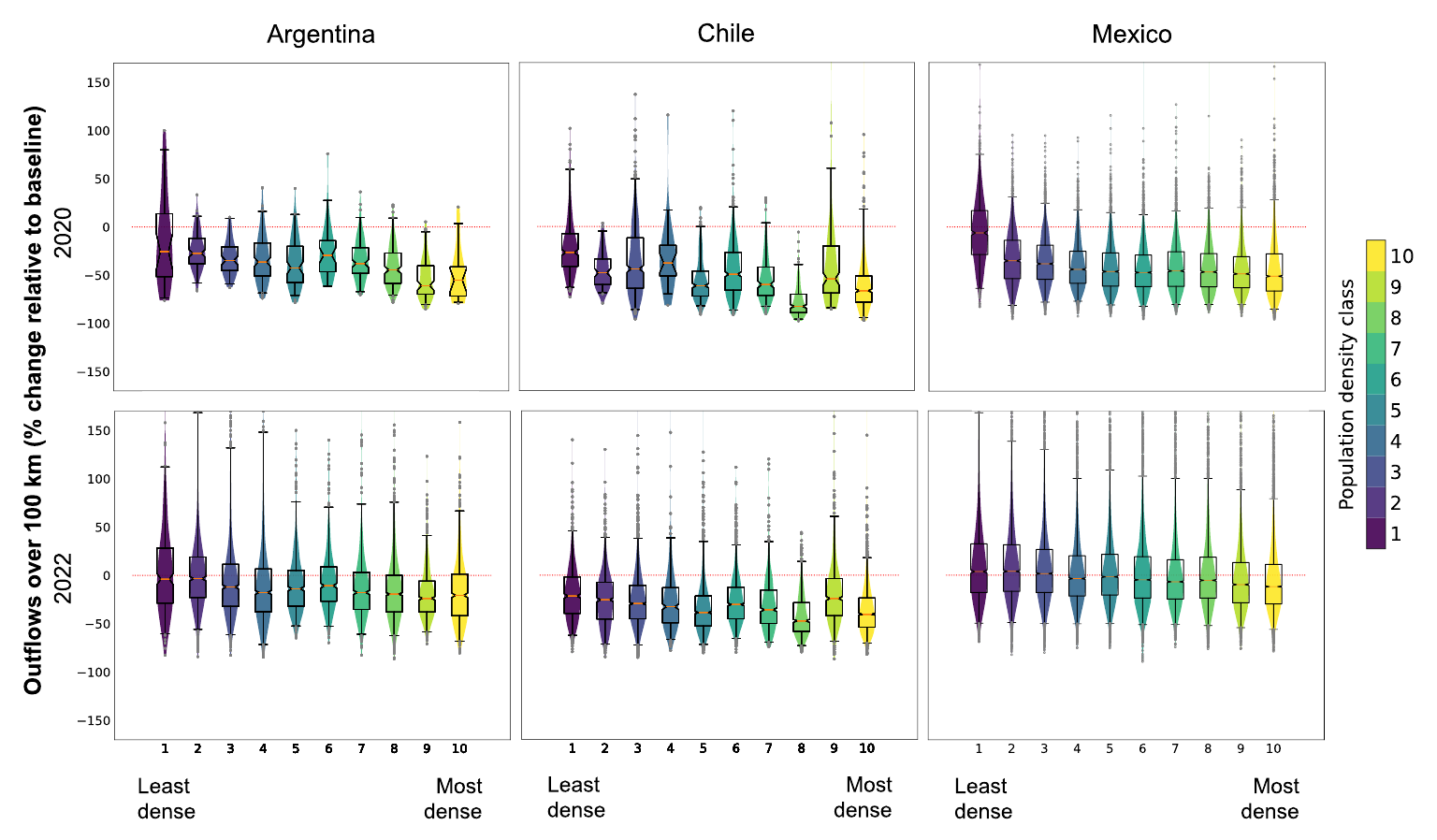}

}

\caption{\label{fig-outflowsO100}Changes in mobility flows during May
2020 and March 2022 by population density class, relative to baseline
period. Long-distance (\textless100km) movements.}

\end{figure}

We analysed changes in the intensity of movements across our population
density classes across Argentina, Chile and Mexico in May 2020 and March
2022. These two months represent two pivotal points in the pandemic. May
2020 provides a good representation of the early days when a series of
strong stringency measures were enacted following the WHO's declaration
of COVID-19 on the 11th of March 2020 as a global pandemic. March 2022
captures the later days of the COVID-19 pandemic about six months after
most of the COVID-19 restrictions had been relaxed in the countries in
our analysis. Figure~\ref{fig-outflowsU100} and
Figure~\ref{fig-outflowsO100} show boxplots of the distribution of the
percentage change in the number of movements by areas according to their
population density for under and over 100km. The boxplots report the
percentage change in movements for each population density category
during May 2020 and March 2022. The baseline levels are represented by
the dotted line at \(y=0\). Positive values indicate increases in
mobility relative to pre-pandemic levels, while negative values indicate
a reduction in mobility.

\emph{Short-distance movements}

Figure~\ref{fig-outflowsU100} shows a consistent decline in movement
across all population density classes for movements under 100km. This
decline is consistent across all countries in our analysis and
particularly pronounced in the early days of the pandemic.
Figure~\ref{fig-outflowsU100} also shows a consistent gradient of
decline. The most densely populated areas, comprising locations within
Buenos Aires, Santiago and Mexico City, tended to record the sharpest
declines in the early stages of the pandemic. Some of these areas
registered drops of over 50\%, notably in Chile. Declines were less
pronounced in the least densely populated areas.

While Figure~\ref{fig-outflowsU100} shows that the predominant trend in
May 2020 was a decline in the levels of mobility, it also reveals that
there is a great degree of variability. While most areas experienced a
decline in May 2020, others recorded larger volumes of movement in
relation to the prepandemic period, and this occurs predominantly in the
least density class areas. Some doubled their levels of mobility. This,
however, does not necessarily mean large population numbers as least
density class areas house very small populations, and small variations
in population translate into large percentage changes. Yet, these
percentage change provide an idea of the degree of impacts on these
sparsely populated locations.

Additionally, Figure~\ref{fig-outflowsU100} reveals that an overall
recovery of mobility levels. While some areas continued to record levels
of mobility well below their prepandemic patterns,
Figure~\ref{fig-outflowsU100} shows that average levels of mobility have
moved closer to prepandemic levels in most areas in March 2022 with
Chile displaying the most modest intensity of recovery. This evidence
suggests that the pandemic may have resulted in durable changes in the
patterns of short-distance mobility, particularly in highly dense areas.
Such change may be the result of the adoption of remote or hybrid
working as a main way of operation.

\emph{Long-distance movements}

Figure~\ref{fig-outflowsO100} displays similar trends for long-distance
movements to those observed for short-distance trips. It shows a
consistent decline in movement across all population density classes in
May 2020, with the most densely populated areas experiencing the largest
drops, and small average changes in the least dense areas. Similarly,
Figure~\ref{fig-outflowsO100} shows a large degree of variability within
population density classes, yet the extent of increases was less
pronounced than for short-distance movements. These findings indicate
that the enactment of stringency measures was very effective in reducing
mobility levels and potentially the extent of COVID-19 transmission in
the early days of the pandemic.

Figure~\ref{fig-outflowsO100} also reveals that the average intensity of
long-distance movement were virtually back to prepandemic levels in
Argentina and Mexico. This trend is evidence by the boxes overlapping
the baseline 0. However, Figure~\ref{fig-outflowsO100} suggests that the
mobility levels were still below prepandemic volumes in Chile,
particularly in moderate and high population density areas. Our
estimates indicate that long-distance mobility levels were 50\% below
prepandemic in the highest density areas covering the Santiago
metropolitan area.

\hypertarget{spatio-temporal-patterns-of-population-redistribution-during-covid-19}{%
\subsection{Spatio-temporal patterns of population redistribution during
COVID-19}\label{spatio-temporal-patterns-of-population-redistribution-during-covid-19}}

\begin{figure}

{\centering \includegraphics{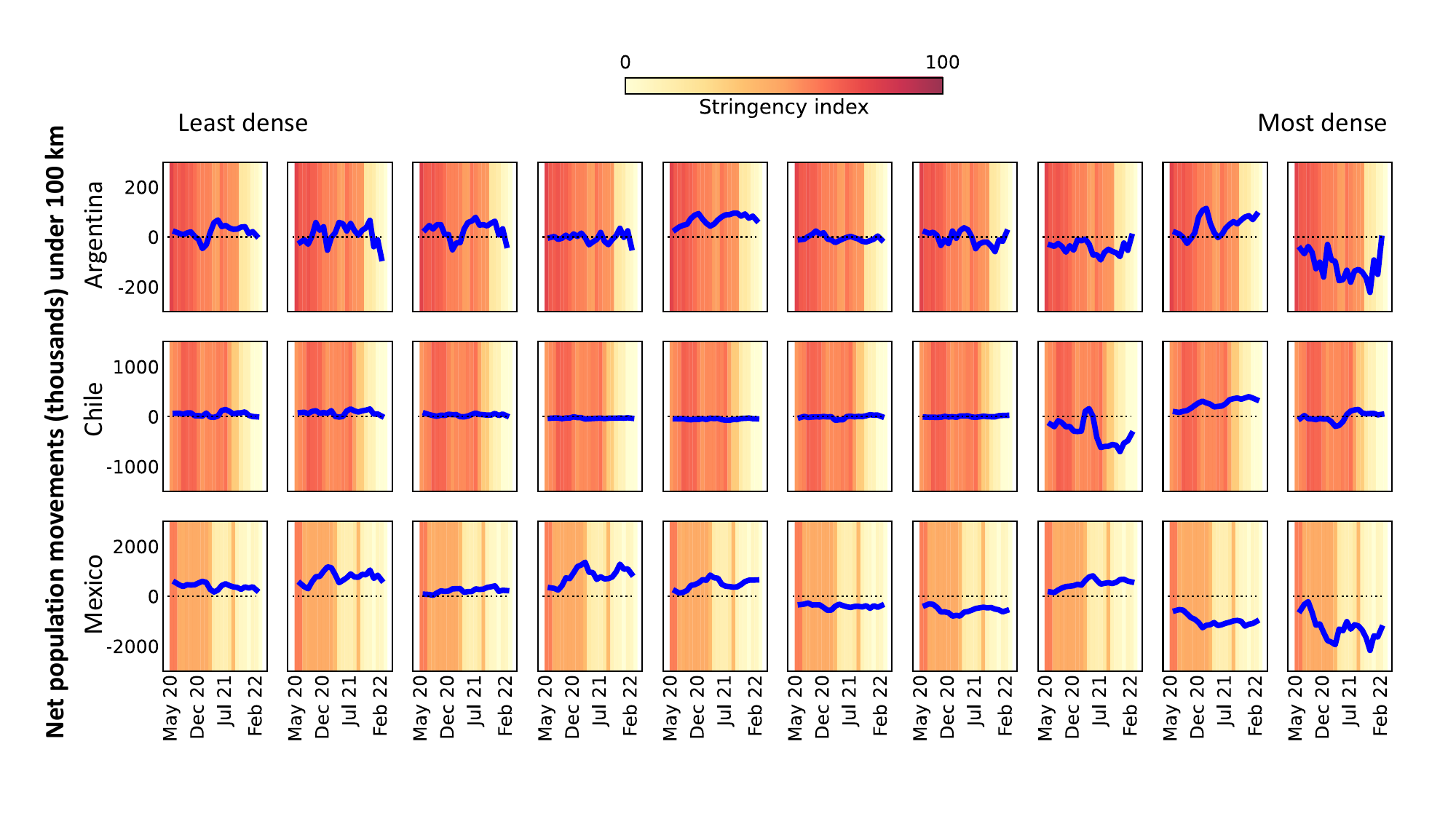}

}

\caption{\label{fig-fig4}Total number of net movement by population
density class. Short-distance (\textless100km) movements.}

\end{figure}

\begin{figure}

{\centering \includegraphics{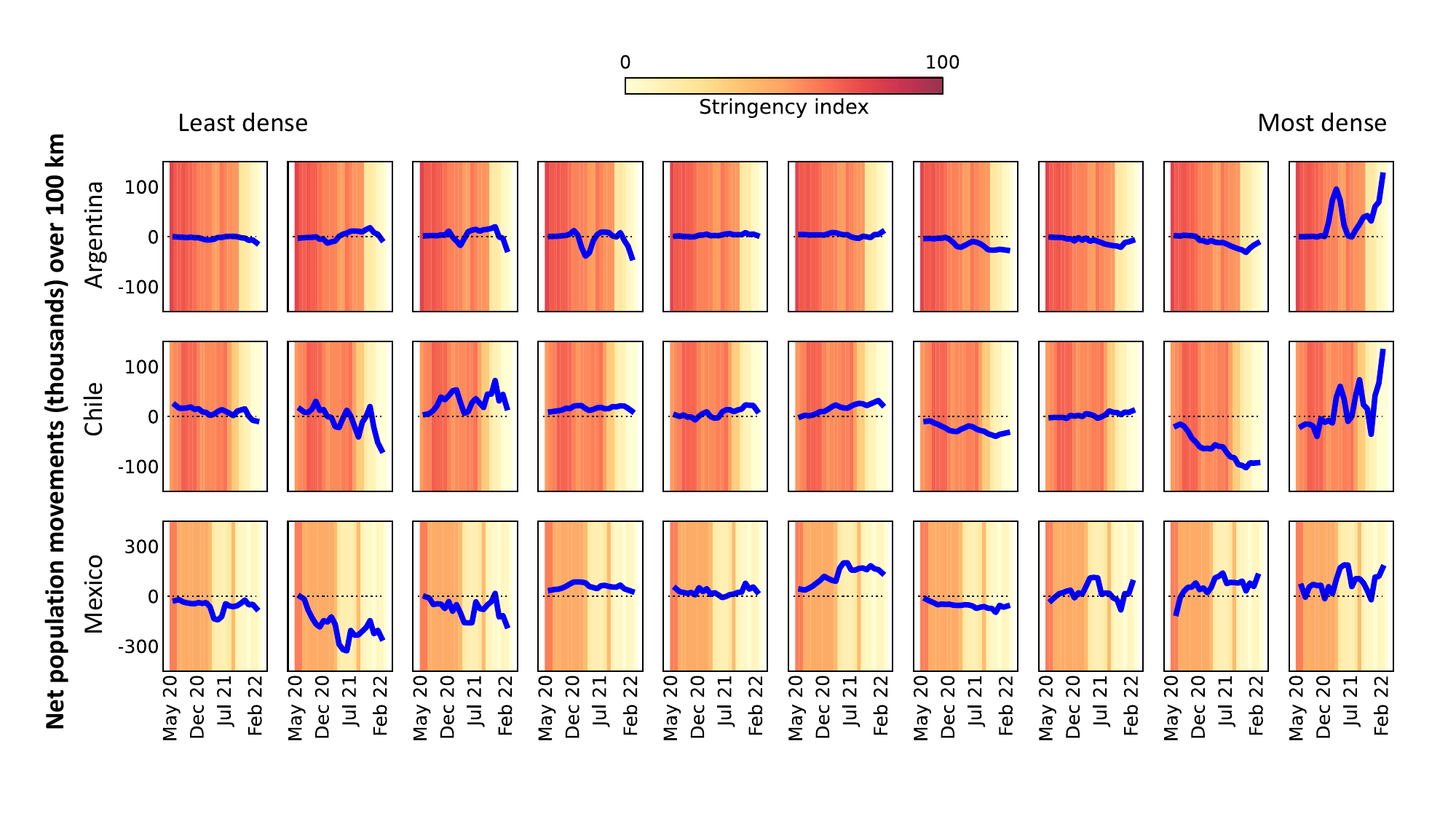}

}

\caption{\label{fig-fig5}Total number of net movement by population
density class. Long-distance (\textgreater100km) movements.}

\end{figure}

We also analysed the spatial impact of mobility on redistributing
population across the country during the COVID-19 pandemic.
Specifically, we examined the evolution of changes in the net balance
between mobility inflows and outflows across the rural-urban hierarchy
during the pandemic. We calculated the monthly net movement balance as
the difference between mobility inflows minus outflows for individual
population density classes during March 2020 to May 2022.
Figure~\ref{fig-fig4} and Figure~\ref{fig-fig5} show the changes in net
balance for movements under 100km and over 100km, respectively.

\emph{Short-distance movement}

Figure~\ref{fig-fig4} reveals a persistent pattern of fluctuating
negative net balances of movements under 100km in the highest population
density class areas in Argentina and Mexico for most of March 2020 to
May 2022. The extent of these balances becomes more notably pronounced
after July 2020, but they vary widely by country. Mexico displays the
largest negative balances, probably reflecting the fact that it is the
second most populated country in Latin America following Brazil. The
trend of negative balances indicates that the highest population density
class areas in Argentina and Mexico tended to record a larger number of
outward movements than inward movements over distances of less than
100km. As indicated in Section~\ref{sec-methods1}, these locations
comprise highly dense metropolitan cores predominantly in capital
cities, including the Central Business District. In Mexico, consistently
large negative net balances also occurred in areas of class 9. These
areas include citie such as Monterrey, Puebla and Guadalajara. The
spatial concentration of negative net balances in high density class
areas mirrors the pattern of population loss identified in metropolitan
areas in developed countries during the COVID-19 pandemic (Rowe,
González-Leonardo, and Champion 2023).

At the same time, Figure~\ref{fig-fig4} shows a relatively consistent
positive net balance of movements in specific types of areas along the
rural-urban hierarchy. Argentina records a consistent pattern of greater
inward movement than outward movement in medium density areas of class 5
type, including small towns and cities such as Ushuaia, Trelew and San
Luis, and a trend of irregular net positive balances in the least dense
areas of class 1-3 and high density locations of class 9. Areas of class
1-3 in Argentina correspond to rural and sparsely populated regions
across the country, while class 9 areas predominantly represent suburban
metropolitan and large city areas, such as Cordoba, Mendoza and Moreno.
Negative net losses also tended to occur in areas of class 7 and 8 over
2021 and 2022, predominantly including medium-sized cities such as Salta
and Resistencia. In Mexico, a pattern of consistently positive net
balance of movements occurred in the range of less dense areas of class
1 to 5 and 8. These balances were particularly pronounced in areas of
class 2 and 4, encompassing touristic, forest and natural reserve places
such as Rio Verde, Chilcota and Ciudad Valles. In Mexico, the observed
patterns suggest mobility out of high density areas in Mexico City,
including metropolitan core areas and suburbs, to nearby sparsely and
rural areas within a radius of 100km.

Chile displays a different pattern. The most densely populated areas in
the country record net balances around zero for movements under 100km.
This pattern suggests limited population redistribution from and to
metropolitan core areas in Santiago over short distances during the
pandemic. Yet, areas of class 8 display large net losses, particularly
from July 2021 following less stringent COVID-19 restrictions. At the
same time, areas of class 9 report more pronounced net gains since July
2021. These findings suggest a pattern of movement up the rural-urban
hierarchy from urban areas, such as Valpara\'iso, La Serena and Viña del
Mar areas to more dense locations in the Santiago Metropolitan Area.

\emph{Long-distance movements}

Figure~\ref{fig-fig5} reveals a systematic pattern of net balances
across Argentina, Chile and Mexico. It shows small net balances of
movements over 100km in the most densely populated areas from the start
of the COVID-19 outbreak to December 2020 but sizable net gains during
2021 and 2022. The pace of these gains, however, differs across
countries. Argentina and Chile report rapid a rise in net gains of
long-distance movements in metropolitan core areas, while the trend of
increase in positive balances in Mexican highly dense areas is less
marked. These patterns seem to relate to the impact of stringency levels
of COVID-19 restrictions. These impacts are particularly visible in
metropolitan core areas. Limited population redistribution appears to
have occurred early in the pandemic when severe COVID-19 restrictions
were enacted, and enlarged as the stringency level of these measures
decreased. Also, comparatively, Argentina and Chile implemented stricter
restrictions and reductions in stringency levels seem to have resulted
in larger expansions in the net balances of metropolitan core areas than
in Mexico.

Unlike other countries, Chile displays a pattern of negative net
balances of movements over 100km in the highest density areas of the
Santiago metropolitan area during the first year of the pandemic
(Figure~\ref{fig-fig5}). These balances are not very large but suggest a
steady flow of people moving away from high density areas of Santiago to
destinations over 100km. Net positive balances of movements over 100km
in less dense areas of class 1 to 3 indicate that these were the primary
destinations of flows from the capital. Yet, the duration of these
balances suggests that movements from metropolitan core areas were
temporary. Figure~\ref{fig-fig5} reveals an irregular pattern of large
net positive balances in these areas from May 2021 as COVID-19
restrictions were relaxed. At the same time, we observe a continuous
trend of negative net balances in highly dense areas of class 9,
reflecting increasing net losses in north-west areas of the Santiago
Metropolitan Area, including Quilicura and Lampa, as well as Reñaca.

\hypertarget{movements-from-and-to-capital-cities}{%
\subsection{Movements from and to capital
cities}\label{movements-from-and-to-capital-cities}}

\begin{figure}

{\centering \includegraphics{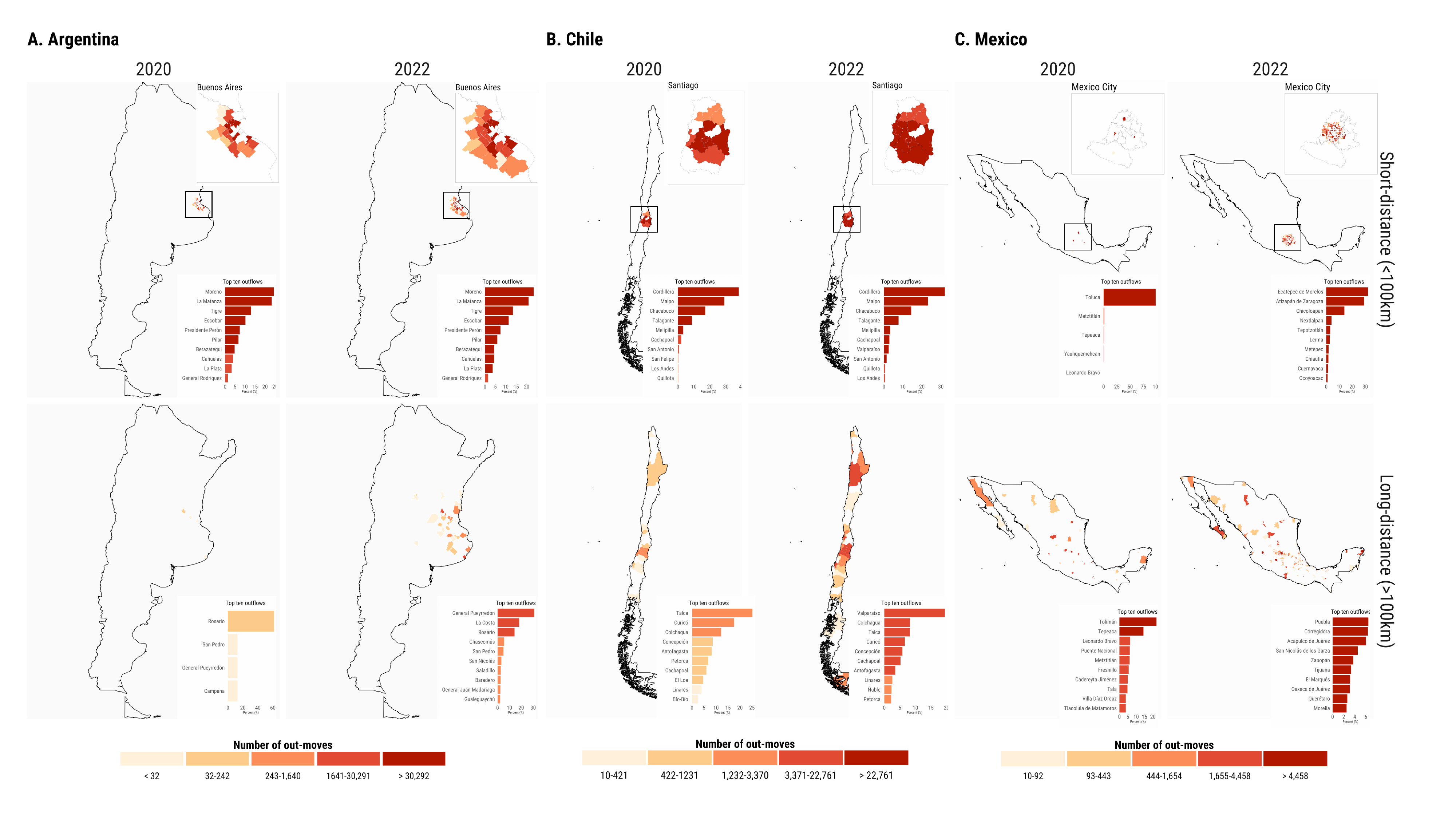}

}

\caption{\label{fig-fig6}Short-distance (\textless100km) and
long-distance (\textgreater100km) movements from core areas in capital
cities. April 2020 and March 2022.}

\end{figure}

To understand better the nature of movements involving areas of
population density class 10, we analysed the spatial patterns of
movement of under and over 100km from capital cities. In our analysis,
movements from population density class 10 areas exclusively represent
mobility patterns from metropolitan core areas of capital cities.
Figure~\ref{fig-fig6} displays the number of movements out from
metropolitan core areas in Buenos Aires, Santiago and Mexico City for
March 2020 and April 2022. As anticipated, areas around the national
capitals represent the main destinations for short-distance movements.
In April 2020, movements away from capital were limited but had expanded
in volume and geographical extent by March 2022.

Figure~\ref{fig-fig6} reveals the main destinations for the movements
underpinning the persistent losses of short-distance movement observed
for class 10 areas in Argentina and Mexico. Our data revealed that
nearby urban and suburban areas, such as Moreno, La Matanza and Tigre in
Argentina and Ecatepec de Morelos, Aztizapán de Zaragoza and Chicoloapan
de Juárez in Mexico absorbed over 80\% of short-distance moves from
Buenos Aires and Mexico City in March 2022. To a lesser extent, these
moves have been directed to rural areas. Though, as we described in the
previous section, these areas recorded a relatively large net balance of
short-distance moves particularly in Mexico. In Chile, short-distance
movements were quite balanced in metropolitan core areas throughout from
2020 and 2022 (Figure~\ref{fig-fig4}). Figure~\ref{fig-fig6} reveals
that these movements were predominantly to mostly rural, sparsely
populated areas, such as Cordillera and Maipo.

Figure~\ref{fig-fig6} also reveals a similar pattern to the donut effect
documented in the US with large cities hollowing out (Ramani and Bloom
2021). The maps for short-distance moves in Figure~\ref{fig-fig6} show
that suburban areas closer to metropolitan core areas absorbed larger
counts of short-distance moves, with neighbouring rural locations and
towns attracting smaller numbers. In Argentina and Chile, this pattern
seem to have been relatively uniform with areas closer to core
metropolitan areas of Buenos Aires and Santiago absorbing large numbers.
In Mexico, areas on the west and north-west, such as Lerma, Metepec and
Chiautla have tended to record the largest numbers from Mexico City.

Figure~\ref{fig-fig6} identifies the destination areas underpinning the
negative balance of long-distance movements in the early months of the
pandemic observed for Santiago, Chile. The 2020 bottom map for
long-distance in Figure~\ref{fig-fig6} reveal that key destinations in
Chile were central-southern areas in the provinces of Talca, Curico and
Colchagua. Movements to these destinations are thought to have been
moves from affluent households, with second or holiday homes in these
areas and jobs which could be done remotely. Additionally,
Figure~\ref{fig-fig6} shows that the provinces of Antofagasta and El Loa
also attracted a relatively large number of moves. The main economic
sector in these areas is mining suggesting that movements to these
regions may have been long-distance commuting moves, rather than
migratory moves.

\begin{figure}

{\centering \includegraphics{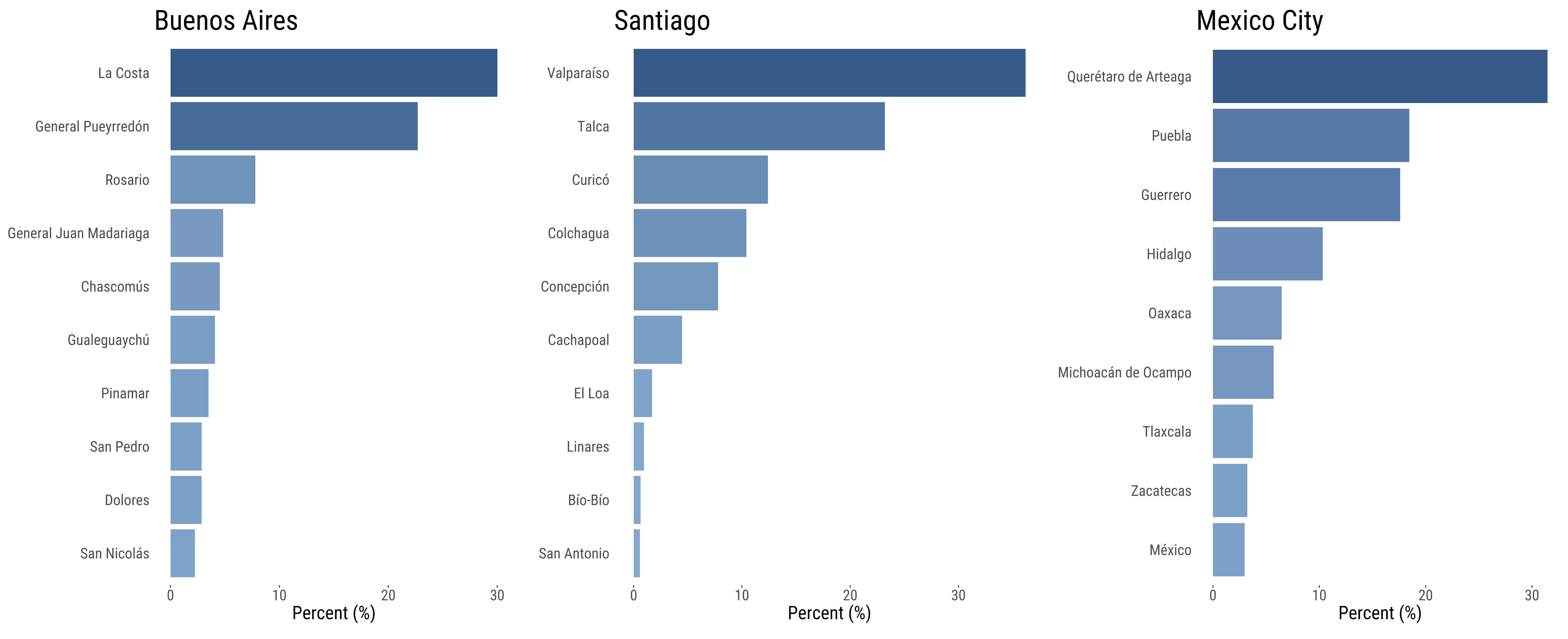}

}

\caption{\label{fig-fig7}Long-distance (\textgreater100km) movements to
metropolitan core areas in capital cities. March 2022. Top ten largest
flows.}

\end{figure}

Figure~\ref{fig-fig5} showed that metropolitan core areas in the three
countries in our analysis have tended to record consistent positive
balances of long-distance movements from early 2021.
Figure~\ref{fig-fig7} reports the top ten areas of origins for these
flows in March 2020, revealing that over 25\% of all inflows over 100km
were originated from coastal areas in Argentina and Chile and from the
small central state of Querétaro in Mexico. These are known as popular
tourism destination suggesting that these flows may reflect temporary
moves made to second or holiday homes by capital city residents during
the pandemic. Contrary to early speculations of the death of cities,
this pattern also highlights the resilience of cities to endure crises
and as a key engine of economic growth, innovation and prosperity,
attracting population from long distances.

\hypertarget{sec-discussion}{%
\section{Discussion}\label{sec-discussion}}

\hypertarget{key-findings}{%
\subsection{Key findings}\label{key-findings}}

Emerging evidence has indicated that the COVID-19 pandemic resulted in a
urban exodus from large cities in developed countries. Anecdotal
evidence has suggested that a similar pattern occurred in Latin American
countries. However, lack of access to suitable data has prevented the
analysis of changes in internal population movements over the pandemic.
Drawing on location data from Meta-Facebook users, we sought to analyse
the extent and durability of changes in internal population movements
across the rural--urban hierarchy in Argentina, Chile and Mexico during
the COVID-19 pandemic from March 2020 to May 2022. We found evidence of
an overall systematic decline in the level of population internal
movement during the enactment of nonpharmaceutical interventions in May
2020 across all three countries. We showed that declines occurred across
the rural-urban continuum for movements over both short and long
distances. The largest reductions occurred in the most dense areas of
Argentina, Chile and Mexico. Average levels of declines in less dense
areas were less pronounced as a large number of them recorded rises
mobility levels. We also revealed that the intensity of mobility bounced
back closer to pre-pandemic levels in May 2022 following the relaxation
of COVID-19 stringency measures. Yet, the intensity of movement remained
below pre-pandemic levels in many areas.

Additionally, we presented evidence of sustained negative net balances
of movements under 100km in the highest density areas of Argentina and
Mexico during 2020 and 2022, particularly since July 2020. At the same
time, we identified a positive net balance of movements for distances
under 100km in areas of low and moderate population density. In Chile,
population redistribution for movements under 100km is limited across
the urban hierarchy, except for moderately high density areas. These
patterns differ from those observed for movements over 100km. In all
three countries, the most dense areas registered overall positive net
balances of movements over 100km particularly from early 2021. Only in
Chile, we found evidence of a negative net balance for movements over
100km in the hightest density areas during 2020, when COVID-19 began
spreading throughout the country and stringency measures were enforced.
The patterns of population redistribution at lower levels of the urban
hierarchy varied by country. In Argentina, we observed moderate negative
net balances of movements over 100km in dense areas. In Chile, a
pronounced and growing negative net balance for these movements occurred
in highly dense areas around the Santiago Metropolitan Area and
neighbouring cities. In Mexico, large negative net balances were
recorded in low density areas.

\hypertarget{interpretation}{%
\subsection{Interpretation}\label{interpretation}}

Our findings suggest that levels of internal population movement over
both short and long distances declined in 2020 and remained below
pre-pandemic levels in May 2022. In Chile, seasonality in mobility
patterns may prompt these conclusions as the pre-pandemic baseline
coincides with the national period of summer holidays when levels of
internal movements over long distances are typically higher than during
the rest of the year. Yet, we argue that such seasonality produces a
substitution effect by trading short-distance commuting moves for
long-distance travel, rather than increasing overall levels of mobility
across the country. Additionally, we argue that hybrid forms of working
that emerged during the height of the pandemic may have had a continuing
effect on internal population movements in the three countries in our
analysis. Hybrid working may have precipitated a secular trend of
internal migration decline which was in motion before the pandemic
(Martin Bell et al. 2017), and led to reduced rates of mobility in many
countries during COVID-19 (Rowe, González-Leonardo, and Champion 2023).
Furthermore, global analyses indicate that hybrid working is thus
expected to reduce the need for daily commuting travel and living in
close proximity to work places (Barrero, Bloom, and Davis 2021; Aksoy et
al. 2022). Hybrid working has become a new future of our working life as
stigma associated with working from home has diminished (Aksoy et al.
2022). Large investments on working environments to enable hybrid work
have been made; and, working-from-home experiences in terms of
productivity and work-life balance have been better than expected
(Barrero, Bloom, and Davis 2021; Aksoy et al. 2022).

Additionally, our study lends some support for the theory of an urban
exodus providing evidence of negative net balances of movements in the
most dense areas of capital cities. These patterns are consistent with
`the donut effect' identified by Ramani and Bloom (2021) to describe net
population losses in dense city centres in most US metropolitan areas
during ealy stages of the COVID-19 pandemic. In Argentina, Chile and
Mexico, distinct processes seem to have underpinned these patterns of
negative net balances. In Argentina, sustained negative balances in
highly dense areas were associated with short-distance movements, likely
reflecting a pattern of suburbanisation in the metropolitan area of
Buenos Aires. In Mexico, the pattern of negative balances were also
associated with short-distance moves but extended to a wider set of
highly dense areas around the metropolitan area of Mexico City, and
generated positive balances of moves in sparsely populated areas,
suggesting a pattern of counterurbanisation. In Chile, negative balances
in highly dense areas of the metropolitan area of Santiago were
underpinned by long-distance moves. These losses were however temporary,
switching to neutral and positive balances in 2021 and 2022.

Overall, we contributed to expanding existing evidence on the impact of
COVID-19 on the levels and spatial patterns of internal population
movement. Existing work has focused on developed countries and
restricted to the immediate effects of COVID-19 analysing data from
2020. The cumulative evidence suggest that the levels of human mobility
declined in 2020, that large cities lost population through internal
migration, and that suburbs and rural areas experienced population
gains. To our knowledge, only two studies have analysed data extending
to 2021 for Spain (González-Leonardo and Rowe 2022) and Britain (Rowe et
al. 2022). In Spain, large cities continued to report population losses
through internal migration. In Britain, low and high density areas
returned to display net balances of movement similar to those observed
before the COVID-19 pandemic.

We expanded the existing evidence by investigating the patterns of
internal population movement in Latin America over a period stretching
from March 2020 to May 2022. Our evidence suggests that negative
balances of moves in highly dense areas and large cities recorded for
various countries in 2020 may have persisted over 2021 and 2022, as we
documented for short-distance moves in Argentina and Mexico. They
suggest a structural change from a core city preference for suburban and
less populated areas. Though, this seemed to be shifting back in
Argentina. Alternatively, our findings suggest that population losses
due to long-distance movement in highly dense areas in the metropolitan
area of Santiago were short-lived, shifting to neutral and positive
balances over 2021 and 2022. Contrary to the initial expectations of
apocalyptic headlines - `empty cities', `dead cities', `ghost cities',
`deserted cities' and `silent cities'- on the media, the evidence
emerging from Chile attests the attraction of cities to absorb
population. Economies agglomeration in dense urban spaces are likely to
continue to facilitate and foster knowledge exchange, innovation and
economic growth (Storper and Venables 2004). At the same time, digital
infrastructure in rural locales of Latin American countries remain
deficient (Pick, Sarkar, and Parrish 2020), representing a major
challenge to permanently relocate to these areas.

\hypertarget{limitations-and-future-work}{%
\subsection{Limitations and future
work}\label{limitations-and-future-work}}

We distinguished between short- and long-distance moves. Typically
short-distance moves are typically used to capture local residential and
routine moves, and long-distance moves are used to account for more
permanent and infrequent moves, such as internal migration (Owen and
Green 1992). However, we cannot identify moves involving permanent
changes of residential address. Our data capture these movements but
also include frequent and daily travel behaviour, such as commuting and
shopping trips. Future work could extend and validate our work
identifying changes to moves relating permanent and temporary
displacements, and triangulating alternative DFD and traditional data
sources, such as censuses and representative surveys collecting
information on residential location.

We focused on analysing the levels and spatial patterns of internal
population movement. Available evidence from the Global North suggests
systematic differences in the mobility patterns of more and less
socio-economically advantaged individuals during the early stages of the
pandemic (Gauvin et al. 2021; Long and Ren 2022; Santana et al. 2023).
While empirical evidence is lacking, existing findings suggest that more
socio-economically advantaged individuals seem to have been able to
relocate more ``permanently'' away from large and dense urban areas
during the height of the pandemic (Gauvin et al. 2021). They are more
likely to have second or holidays homes and have jobs which can be done
remotely. Less socio-economically advantaged individuals were less able
to relocate away from their home as they are more likely to work on less
well paid service jobs requiring face-to-face interaction (Santana et
al. 2023). Drawing on area-level population attributes, future work
could investigate the occurrence and persistence of these patterns in
Global South countries where socio-economic inequalities are more acute
(Qureshi 2023).

Further research is also needed to establish the cause of the observed
changes in the spatial patterns of internal population movement since
the start of the COVID-19 pandemic. Understanding these causes can
assist in anticipating long-term structural changes in the direction of
mobility flows in the coming years. A combination of factors, including
social distancing, business shutdowns, school closures, telework and
unemployment, have all been cited as key forces recasting preexisting
patterns of population movement in the early stages of the pandemic.
While some of these factors vanished as COVID-19 restrictions were
relaxed, others such as telework has endured the pandemic (Aksoy et al.
2022). Determining the extent to which forms and workers have and will
continue to adopt remote work will be key to understand future
residential and mobility choices within and away from large and dense
urban areas.

Additionally, accessing and working with DFD to measure human mobility
and migration is challenging. DFD are known to suffer from biases and
issues of statistical representation reflecting differences in
technology usage and accessibility (Rowe 2023). While concerted efforts
to develop data services and methodological frameworks to leverage on
DFD exist in the Global North {[}REF{]}, limited progress has been done
in Latin America. National government and statistical agencies have a
prime position to promote the creation of data sharing partnerships with
private companies and create the digital infrastructure to ensure equal
and secure access to DFD sources for the social good. With academia,
these agencies could play a major role in promoting the creation of a
methodological framework for the use and analysis of DFD, setting common
definitions, metrics, and standard approaches to measure and correct
data biases and validate data analyses. Location data have been proven
key to monitor and develop appropriate policy responses to population
displacement and movement during natural disasters, wars and epidemics
(Green, Pollock, and Rowe 2021). We envision resources like those
created by the European Commission Joint Research Centre (2022) and the
UNSD (2019), identifying sources of non-traditional data and
establishing methodological guidelines for the use of mobile phone data
for official mobility statistics. Existing resources have a global
character, we call for initiatives with a regional focus accounting for
differences in the local landscape of data availability and digital
technology usage.

\hypertarget{sec-conclusion}{%
\section{Conclusion}\label{sec-conclusion}}

The COVID‐19 pandemic significantly altered civil liberties and the ways
people move in 2020. Over the last two years, research has been
undertaken to understand the immediate impacts of the COVID‐19 pandemic
on human mobility particularly in the Global North. Limited research has
focused on understanding the persistence of the impacts of the COVID‐19
pandemic beyond 2020 in the Global South context, primarily due to the
lack of suitable data. Drawing on aggregate anonymised mobile phone
data, we examined the extent and durability of changes in the levels and
spatial patterns of internal population movement across the rural-urban
continuum in Argentina, Chile and Mexico over a 26-month period from
March 2020 to May 2022. We presented evidence of an overall systematic
decline in the level of short- and long-distance movement during the
enactment of nonpharmaceutical interventions in 2020, with the largest
reductions occurring in the most dense areas. While levels of movements
recovered and moved closer to pre-pandemic levels in 2022, they have
remained below pre-pandemic levels. This trend aligns with a broader
pattern of declining migration intensities in many industrialised
countries (Martin Bell et al. 2017). Additionally we also showed
evidence offering some support for the idea of an urban exodus. We found
a continuing negative net balance of short-distance movements in the
highly dense areas of capital cities in Argentina and Mexico, reflecting
a pattern of suburbanisation. In Chile, we recorded very limited changes
in the net balance of short-distance movements but we found a net loss
of long-distance movements. However, these losses were temporary. Yet,
our findings point to long-lasting effects of COVID-19 in Argentina and
Mexico, resulting in sustained losses of movements in core metropolitan
areas of capital cities and gains in suburban locations. These patterns
are contrary to population gains through internal migration in cities
over 1 million people in the region of Latin America and Caribbean
(Rodr\'iguez Vignoli 2017), and may reflect the enduring effect of remote
work and differences in housing costs as centrifugal forces.

\hypertarget{references}{%
\section*{References}\label{references}}
\addcontentsline{toc}{section}{References}

\hypertarget{refs}{}
\begin{CSLReferences}{1}{0}
\leavevmode\vadjust pre{\hypertarget{ref-aksoy2022}{}}%
Aksoy, Cevat Giray, Jose Maria Barrero, Nicholas Bloom, Steven Davis,
Mathias Dolls, and Pablo Zarate. 2022. {``Working from Home Around the
World.''} \url{https://doi.org/10.3386/w30446}.

\leavevmode\vadjust pre{\hypertarget{ref-aromuxed2023}{}}%
Arom\'i, Daniel, Mar\'ia Paula Bonel, Julian Cristia, Mart\'in Llada, Juan
Pereira, Xiomara Pulido, and Julieth Santamaria. 2023. {``\#StayAtHome:
Social Distancing Policies and Mobility in Latin America and the
Caribbean.''} \emph{Econom\'ia} 22 (1).
\url{https://doi.org/10.31389/eco.4}.

\leavevmode\vadjust pre{\hypertarget{ref-barrero2021}{}}%
Barrero, Jose Maria, Nicholas Bloom, and Steven Davis. 2021. {``Why
Working from Home Will Stick.''} \url{https://doi.org/10.3386/w28731}.

\leavevmode\vadjust pre{\hypertarget{ref-bell2017global}{}}%
Bell, Martin, Elin Charles-Edwards, Aude Bernard, and Philipp Ueffing.
2017. {``Global Trends in Internal Migration.''} In \emph{Internal
Migration in the Developed World}, edited by T Champion, T Cooke, and I
Shuttleworth, 76--97. Routledge.

\leavevmode\vadjust pre{\hypertarget{ref-bell2002}{}}%
Bell, M, M Blake, P Boyle, O Duke-Williams, P Rees, J Stillwell, and G
Hugo. 2002. {``Cross-National Comparison of Internal Migration: Issues
and Measures.''} \emph{Journal of the Royal Statistical Society: Series
A (Statistics in Society)} 165 (3): 435--64.
\url{https://doi.org/10.1111/1467-985x.00247}.

\leavevmode\vadjust pre{\hypertarget{ref-bernard2017}{}}%
Bernard, Aude, Francisco Rowe, Martin Bell, Philipp Ueffing, and Elin
Charles-Edwards. 2017. {``Comparing Internal Migration Across the
Countries of Latin America: A Multidimensional Approach.''} Edited by
Osman Alimamy Sankoh. \emph{PLOS ONE} 12 (3): e0173895.
\url{https://doi.org/10.1371/journal.pone.0173895}.

\leavevmode\vadjust pre{\hypertarget{ref-borsdorf2003}{}}%
Borsdorf, Axel. 2003. {``Cómo Modelar El Desarrollo y La Dinámica de La
Ciudad Latinoamericana.''} \emph{EURE (Santiago)} 29 (86).
\url{https://doi.org/10.4067/s0250-71612003008600002}.

\leavevmode\vadjust pre{\hypertarget{ref-Brea03}{}}%
Brea, J. 2003. {``Population Dynamics in Latin America.''}
\emph{Population Bulletin} 58: 3--36.
\url{https://population.un.org/wup/Publications/Files/WUP2018-Report.pdf}.

\leavevmode\vadjust pre{\hypertarget{ref-cepal2022}{}}%
CEPAL. 2022. {``Lineamientos Generales Para La Captura de Datos
Censales: Revisión de Métodos Con Miras a La Ronda de Censos de 2020.''}
\emph{Publicaci\'on de Las Naciones Unidas}.
\url{https://repositorio.cepal.org/server/api/core/bitstreams/fa5dc0a5-9003-4ee2-95a0-bc3bd7e531be/content}.

\leavevmode\vadjust pre{\hypertarget{ref-championhugo2004}{}}%
Champion, Tony, and Hugo Graeme. 2004. \emph{New Forms of Urbanization:
Beyond the Urban-Rural Dichotomy}. Routledge.

\leavevmode\vadjust pre{\hypertarget{ref-chuxe1vezgalindo2016}{}}%
Chávez Galindo, Ana Mar\'ia, Jorge Rodr\'iguez Vignoli, Mario Acuña, Jorge
Barquero, Daniel Macadar, José Marcos Pinto da Cunha, and Jaime Sobrino.
2016. {``Migración Interna y Cambios Metropolitanos.''} \emph{Revista
Latinoamericana de Población} 10 (18): 7--41.
\url{https://doi.org/10.31406/relap2016.v10.i1.n18.1}.

\leavevmode\vadjust pre{\hypertarget{ref-elejalde2023}{}}%
Elejalde, Erick, Leo Ferres, V\'ictor Navarro, Loreto Bravo, and Emilio
Zagheni. 2023. {``The Social Stratification of Internal Migration and
Daily Mobility During the COVID-19 Pandemic.''}
\url{https://doi.org/10.48550/ARXIV.2309.11062}.

\leavevmode\vadjust pre{\hypertarget{ref-europeancommission.jointresearchcentre.2022}{}}%
European Commission Joint Research Centre. 2022. \emph{Data innovation
in demography, migration and human mobility.} LU: Publications Office.
\url{https://doi.org/10.2760/027157}.

\leavevmode\vadjust pre{\hypertarget{ref-fielding2021}{}}%
Fielding, Tony, and Yoshitaka Ishikawa. 2021. {``COVID{-}19 and
Migration: A Research Note on the Effects of COVID{-}19 on Internal
Migration Rates and Patterns in Japan.''} \emph{Population, Space and
Place} 27 (6). \url{https://doi.org/10.1002/psp.2499}.

\leavevmode\vadjust pre{\hypertarget{ref-firebaugh1979}{}}%
Firebaugh, Glenn. 1979. {``Structural Determinants of Urbanization in
Asia and Latin America, 1950- 1970.''} \emph{American Sociological
Review} 44 (2): 199. \url{https://doi.org/10.2307/2094505}.

\leavevmode\vadjust pre{\hypertarget{ref-florida2021}{}}%
Florida, Richard, Andrés Rodr\'iguez-Pose, and Michael Storper. 2021.
{``Critical Commentary: Cities in a Post-COVID World.''} \emph{Urban
Studies} 60 (8): 1509--31.
\url{https://doi.org/10.1177/00420980211018072}.

\leavevmode\vadjust pre{\hypertarget{ref-gauvin2021}{}}%
Gauvin, Laetitia, Paolo Bajardi, Emanuele Pepe, Brennan Lake, Filippo
Privitera, and Michele Tizzoni. 2021. {``Socio-Economic Determinants of
Mobility Responses During the First Wave of COVID-19 in Italy: From
Provinces to Neighbourhoods.''} \emph{Journal of The Royal Society
Interface} 18 (181): 20210092.
\url{https://doi.org/10.1098/rsif.2021.0092}.

\leavevmode\vadjust pre{\hypertarget{ref-ghosh2020}{}}%
Ghosh, Somenath, Pallabi Seth, and Harsha Tiwary. 2020. {``How Does
Covid-19 Aggravate the Multidimensional Vulnerability of Slums in India?
A Commentary.''} \emph{Social Sciences \& Humanities Open} 2 (1):
100068. \url{https://doi.org/10.1016/j.ssaho.2020.100068}.

\leavevmode\vadjust pre{\hypertarget{ref-ginsburg2022}{}}%
Ginsburg, Carren, Mark A. Collinson, F. Xavier Gómez-Olivé, Sadson
Harawa, Chantel F. Pheiffer, and Michael J. White. 2022. {``The Impact
of COVID-19 on a Cohort of Origin Residents and Internal Migrants from
South Africa's Rural Northeast.''} \emph{SSM - Population Health} 17
(March): 101049. \url{https://doi.org/10.1016/j.ssmph.2022.101049}.

\leavevmode\vadjust pre{\hypertarget{ref-gonzuxe1lezollino2006}{}}%
González Ollino, Daniela, and Jorge Rodr\'iguez Vignoli. 2006. {``Spatial
Redistribution and Internal Migration of the Populationin Chile over the
Past 35 Years (1965-20).''} \emph{Estudios Demográficos y Urbanos} 21
(2): 369. \url{https://doi.org/10.24201/edu.v21i2.1253}.

\leavevmode\vadjust pre{\hypertarget{ref-gonzuxe1lez-leonardo2022b}{}}%
González-Leonardo, Miguel, Antonio López-Gay, Niall Newsham, Joaqu\'in
Recaño, and Francisco Rowe. 2022. {``Understanding Patterns of Internal
Migration During the COVID{-}19 Pandemic in Spain.''} \emph{Population,
Space and Place} 28 (6). \url{https://doi.org/10.1002/psp.2578}.

\leavevmode\vadjust pre{\hypertarget{ref-gonzuxe1lez-leonardo2022c}{}}%
González-Leonardo, Miguel, and Francisco Rowe. 2022. {``Visualizing
Internal and International Migration in the Spanish Provinces During the
COVID-19 Pandemic.''} \emph{Regional Studies, Regional Science} 9 (1):
600--602. \url{https://doi.org/10.1080/21681376.2022.2125824}.

\leavevmode\vadjust pre{\hypertarget{ref-gonzuxe1lez-leonardo2022a}{}}%
González-Leonardo, Miguel, Francisco Rowe, and Alberto
Fresolone-Caparrós. 2022. {``Rural Revival? The Rise in Internal
Migration to Rural Areas During the COVID-19 Pandemic. Who Moved and
Where?''} \emph{Journal of Rural Studies} 96 (December): 332--42.
\url{https://doi.org/10.1016/j.jrurstud.2022.11.006}.

\leavevmode\vadjust pre{\hypertarget{ref-gonzuxe1lez-leonardo2023a}{}}%
González-Leonardo, Miguel, Francisco Rowe, and Arturo Vegas-Sánchez.
2023. {``A {`}Donut Effect{'}? Assessing Housing Transactions During
COVID-19 Across the Spanish Rural{\textendash}urban Hierarchy.''}
\emph{Regional Studies, Regional Science} 10 (1): 471--73.
\url{https://doi.org/10.1080/21681376.2023.2191684}.

\leavevmode\vadjust pre{\hypertarget{ref-graizbord2007}{}}%
Graizbord, Boris, and Beatriz Acuña. 2007. {``Movilidad Residencial En
La Ciudad de México / Residential Mobility in Mexico City.''}
\emph{Estudios Demográficos y Urbanos} 22 (2): 291.
\url{https://doi.org/10.24201/edu.v22i2.1281}.

\leavevmode\vadjust pre{\hypertarget{ref-green2021}{}}%
Green, Mark, Frances Darlington Pollock, and Francisco Rowe. 2021.
{``New Forms of Data and New Forms of Opportunities to Monitor and
Tackle a Pandemic.''} In, 423--29. Springer International Publishing.
\url{https://doi.org/10.1007/978-3-030-70179-6_56}.

\leavevmode\vadjust pre{\hypertarget{ref-Hale2021}{}}%
Hale, Thomas, Noam Angrist, Rafael Goldszmidt, Beatriz Kira, Anna
Petherick, Toby Phillips, Samuel Webster, et al. 2021. {``A Global Panel
Database of Pandemic Policies (Oxford COVID-19 Government Response
Tracker).''} \emph{Nature Human Behaviour} 5 (4): 529--38.
\url{https://doi.org/10.1038/s41562-021-01079-8}.

\leavevmode\vadjust pre{\hypertarget{ref-haslag2021}{}}%
Haslag, Peter H., and Daniel Weagley. 2021. {``From L.A. To Boise: How
Migration Has Changed During the COVID-19 Pandemic.''} \emph{SSRN
Electronic Journal}. \url{https://doi.org/10.2139/ssrn.3808326}.

\leavevmode\vadjust pre{\hypertarget{ref-irudayarajan2020}{}}%
Irudaya Rajan, S., P. Sivakumar, and Aditya Srinivasan. 2020. {``The
COVID-19 Pandemic and Internal Labour Migration in India: A {`}Crisis of
Mobility{'}.''} \emph{The Indian Journal of Labour Economics} 63 (4):
1021--39. \url{https://doi.org/10.1007/s41027-020-00293-8}.

\leavevmode\vadjust pre{\hypertarget{ref-janoschka2002}{}}%
Janoschka, Michael. 2002. {``El Nuevo Modelo de La Ciudad
Latinoamericana: Fragmentación y Privatización.''} \emph{EURE
(Santiago)} 28 (85).
\url{https://doi.org/10.4067/s0250-71612002008500002}.

\leavevmode\vadjust pre{\hypertarget{ref-jenks1967data}{}}%
Jenks, George F. 1967. {``The Data Model Concept in Statistical
Mapping.''} \emph{International Yearbook of Cartography} 7: 186--90.

\leavevmode\vadjust pre{\hypertarget{ref-kotsubo2022}{}}%
Kotsubo, Masaki, and Tomoki Nakaya. 2022. {``Trends in Internal
Migration in Japan, 2012{\textendash}2020: The Impact of the COVID{-}19
Pandemic.''} \emph{Population, Space and Place} 29 (4).
\url{https://doi.org/10.1002/psp.2634}.

\leavevmode\vadjust pre{\hypertarget{ref-lattes95}{}}%
Lattes, A. 1995. {``Urbanización, Crecimiento Urbano y Migraciones En
América Latina. Población y Desarrollo: Tendencias y Desaf\'ios.''}
\emph{Notas de Población} 62: 211--60.
\url{https://repositorio.cepal.org/handle/11362/38594}.

\leavevmode\vadjust pre{\hypertarget{ref-lattes2017}{}}%
Lattes, Alfredo E., Jorge Rodr\'iguez, and Miguel Villa. 2017.
{``Population Dynamics and Urbanization in Latin America: Concepts and
Data Limitations.''} In, 89--111. Routledge.
\url{https://doi.org/10.4324/9781315248073-5}.

\leavevmode\vadjust pre{\hypertarget{ref-long2022}{}}%
Long, Jed A., and Chang Ren. 2022. {``Associations Between Mobility and
Socio-Economic Indicators Vary Across the Timeline of the Covid-19
Pandemic.''} \emph{Computers, Environment and Urban Systems} 91
(January): 101710.
\url{https://doi.org/10.1016/j.compenvurbsys.2021.101710}.

\leavevmode\vadjust pre{\hypertarget{ref-lucchini2023}{}}%
Lucchini, Lorenzo, Ollin Langle-Chimal, Lorenzo Candeago, Lucio Melito,
Alex Chunet, Aleister Montfort, Bruno Lepri, Nancy Lozano-Gracia, and
Samuel P. Fraiberger. 2023. {``Socioeconomic Disparities in Mobility
Behavior During the COVID-19 Pandemic in Developing Countries.''}
\url{https://doi.org/10.48550/ARXIV.2305.06888}.

\leavevmode\vadjust pre{\hypertarget{ref-Maas19}{}}%
Maas, P., S. Iyer, A. Gros, W. Park, L. McGorman, C. Nayak, and P. A.
Dow. 2019. {``Facebook Disaster Maps: Aggregate Insights for Crisis
Response and Recovery.''} In \emph{16th International Conference on
Information Systems for Crisis Response and Management}, 836--47.

\leavevmode\vadjust pre{\hypertarget{ref-marsh2020escape}{}}%
Marsh, Sarah. 2020. {``Escape to the Country: How Covid Is Driving an
Exodus from Britain's Cities.''} \emph{The Guardian} 26: 2020.

\leavevmode\vadjust pre{\hypertarget{ref-microsoftBingMaps}{}}%
Microsoft. {``{B}ing {M}aps {T}ile {S}ystem - {B}ing {M}aps.''}
\url{https://learn.microsoft.com/en-us/bingmaps/articles/bing-maps-tile-system}.

\leavevmode\vadjust pre{\hypertarget{ref-nathan2020}{}}%
Nathan, Max, and Henry Overman. 2020. {``Will Coronavirus Cause a Big
City Exodus?''} \emph{Environment and Planning B: Urban Analytics and
City Science} 47 (9): 1537--42.
\url{https://doi.org/10.1177/2399808320971910}.

\leavevmode\vadjust pre{\hypertarget{ref-UNpopulation19}{}}%
Nations", "United. 2019. {``World Urbanization Prospects. The 2018
Revision.''}
\url{https://population.un.org/wup/Publications/Files/WUP2018-Report.pdf}.

\leavevmode\vadjust pre{\hypertarget{ref-niedomysl2011}{}}%
Niedomysl, Thomas. 2011. {``How Migration Motives Change over Migration
Distance: Evidence on Variation Across Socio-Economic and Demographic
Groups.''} \emph{Regional Studies} 45 (6): 843--55.
\url{https://doi.org/10.1080/00343401003614266}.

\leavevmode\vadjust pre{\hypertarget{ref-nouvellet2021}{}}%
Nouvellet, Pierre, Sangeeta Bhatia, Anne Cori, Kylie E. C. Ainslie, Marc
Baguelin, Samir Bhatt, Adhiratha Boonyasiri, et al. 2021. {``Reduction
in Mobility and COVID-19 Transmission.''} \emph{Nature Communications}
12 (1). \url{https://doi.org/10.1038/s41467-021-21358-2}.

\leavevmode\vadjust pre{\hypertarget{ref-owen1992migration}{}}%
Owen, D, and A Green. 1992. {``Migration Patterns and Trends.''} In
\emph{Migration Processes and Patterns: Research Progress and
Prospects}, edited by T Champion and T Fielding. Belhaven Press.

\leavevmode\vadjust pre{\hypertarget{ref-paybarah2020new}{}}%
Paybarah, Azi, Matthew Bloch, and Scott Reinhard. 2020. {``Where New
Yorkers Moved to Escape Coronavirus.''} \emph{The New York Times} 16.

\leavevmode\vadjust pre{\hypertarget{ref-perales2022}{}}%
Perales, Francisco, and Aude Bernard. 2022. {``Continuity or Change? How
the Onset of COVID{-}19 Affected Internal Migration in Australia.''}
\emph{Population, Space and Place} 29 (2).
\url{https://doi.org/10.1002/psp.2626}.

\leavevmode\vadjust pre{\hypertarget{ref-Puxe9rez-Campuzano13}{}}%
Pérez-Campuzano, C., E. y Santos-Cerquera. 2013. {``Tendencias Recientes
de La Migración Interna En México.''} \emph{"Papeles de Población"} 19:
53--88.
\url{https://www.scielo.org.mx/scielo.php?script=sci_arttext\&pid=S1405-74252013000200003}.

\leavevmode\vadjust pre{\hypertarget{ref-pick2020}{}}%
Pick, James, Avijit Sarkar, and Elizabeth Parrish. 2020. {``The Latin
American and Caribbean Digital Divide: A Geospatial and Multivariate
Analysis.''} \emph{Information Technology for Development} 27 (2):
235--62. \url{https://doi.org/10.1080/02681102.2020.1805398}.

\leavevmode\vadjust pre{\hypertarget{ref-PintoDaCunha02}{}}%
Pinto da Cunha, J. M. P. 2002. {``Urbanización, Redistribución Espacial
de La Población y Transformaciones Socioeconómicas En América Latina.''}
\emph{United Nations}.
\url{https://repositorio.cepal.org/bitstream/handle/11362/7168/1/S029663_es.pdf}.

\leavevmode\vadjust pre{\hypertarget{ref-Pomeroy2021}{}}%
Pomeroy, Robin, and Ross Chiney. 2021. {``Has COVID Killed Our Cities?
World Economic Forum.''}
\url{https://www.weforum.org/agenda/2020/11/cities-podcast-new-york-dead/}.

\leavevmode\vadjust pre{\hypertarget{ref-qureshi2023rising}{}}%
Qureshi, Zia. 2023. {``Rising Inequality: A Major Issue of Our Time.''}

\leavevmode\vadjust pre{\hypertarget{ref-ramani2021}{}}%
Ramani, Arjun, and Nicholas Bloom. 2021. {``The Donut Effect of Covid-19
on Cities.''} \url{https://doi.org/10.3386/w28876}.

\leavevmode\vadjust pre{\hypertarget{ref-rodruxedguezvignoli2017}{}}%
Rodr\'iguez Vignoli, Jorge, and Francisco Rowe. 2017. {``¿Contribuye La
Migración Interna a Reducir La Segregación Residencial?''} \emph{Revista
Latinoamericana de Población} 11 (21): 7--45.
\url{https://doi.org/10.31406/relap2017.v11.i2.n21.1}.

\leavevmode\vadjust pre{\hypertarget{ref-rodruxedguez-vignoli2018}{}}%
Rodr\'iguez-Vignoli, Jorge, and Francisco Rowe. 2018. {``How Is Internal
Migration Reshaping Metropolitan Populations in Latin America? A New
Method and New Evidence.''} \emph{Population Studies} 72 (2): 253--73.
\url{https://doi.org/10.1080/00324728.2017.1416155}.

\leavevmode\vadjust pre{\hypertarget{ref-rodriguez2017migracion}{}}%
Rodr\'iguez Vignoli, Jorge. 2017. {``Migraci{ó}n Interna y Asentamientos
Humanos En Am{é}rica Latina y El Caribe (1990-2010).''}
\url{https://repositorio.cepal.org/items/87177405-76af-475e-a730-118841674013}.

\leavevmode\vadjust pre{\hypertarget{ref-rodriguezrowe2019}{}}%
---------. 2019. {``Efectos Cambiantes de La Migraci{ó}n Sobre El
Crecimiento, La Estructura Demogr{á}fica y La Segregaci{ó}n Residencial
En Ciudades Grandes: El Caso de Santiago, Chile, 1977-2017.''}
\emph{Poblaci{ó}n y Desarrollo}.
\url{https://hdl.handle.net/11362/44367}.

\leavevmode\vadjust pre{\hypertarget{ref-rowe2023big}{}}%
Rowe, Francisco. 2023. {``Big Data.''} In \emph{Concise Encyclopedia of
Human Geography}, 42--47. Edward Elgar Publishing.

\leavevmode\vadjust pre{\hypertarget{ref-rowegonzalez}{}}%
---------. 2013. {``Spatial Labour Mobility in a Transition Economy:
Migration and Commuting in Chile.''} PhD thesis.
\url{https://doi.org/10.14264/uql.2017.427}.

\leavevmode\vadjust pre{\hypertarget{ref-rowe2019}{}}%
Rowe, Francisco, Martin Bell, Aude Bernard, Elin Charles-Edwards, and
Philipp Ueffing. 2019. {``Impact of Internal Migration on Population
Redistribution in Europe: Urbanisation, Counterurbanisation or Spatial
Equilibrium?''} \emph{Comparative Population Studies} 44 (November).
\url{https://doi.org/10.12765/cpos-2019-18}.

\leavevmode\vadjust pre{\hypertarget{ref-rowe2022b}{}}%
Rowe, Francisco, Alessia Calafiore, Daniel Arribas-Bel, Krasen
Samardzhiev, and Martin Fleischmann. 2022. {``Urban Exodus?
Understanding Human Mobility in Britain During the COVID{-}19 Pandemic
Using Meta{-}Facebook Data.''} \emph{Population, Space and Place} 29
(1). \url{https://doi.org/10.1002/psp.2637}.

\leavevmode\vadjust pre{\hypertarget{ref-rowe2023}{}}%
Rowe, Francisco, Miguel González-Leonardo, and Tony Champion. 2023.
{``Virtual Special Issue: Internal Migration in Times of COVID{-}19.''}
\emph{Population, Space and Place}, March.
\url{https://doi.org/10.1002/psp.2652}.

\leavevmode\vadjust pre{\hypertarget{ref-rowe2022c}{}}%
Rowe, Francisco, Caitlin Robinson, and Nikos Patias. 2022. {``Sensing
Global Changes in Local Patterns of Energy Consumption in Cities During
the Early Stages of the COVID-19 Pandemic.''} \emph{Cities} 129
(October): 103808. \url{https://doi.org/10.1016/j.cities.2022.103808}.

\leavevmode\vadjust pre{\hypertarget{ref-santana2023}{}}%
Santana, Clodomir, Federico Botta, Hugo Barbosa, Filippo Privitera,
Ronaldo Menezes, and Riccardo Di Clemente. 2023. {``COVID-19 Is Linked
to Changes in the Time{\textendash}space Dimension of Human Mobility.''}
\emph{Nature Human Behaviour}, July.
\url{https://doi.org/10.1038/s41562-023-01660-3}.

\leavevmode\vadjust pre{\hypertarget{ref-sobrino12}{}}%
Sobrino, J. 2012. {``La Urbanización En El México Contemporáneo.''}
\emph{Notas de Población} 94: 93--122.
\url{https://repository.eclac.org/handle/11362/12898}.

\leavevmode\vadjust pre{\hypertarget{ref-sobrino2006}{}}%
Sobrino, Jaime. 2006. {``Comptetitiveness and Employment in the Largest
Metropolitan Areas of Mexico.''} In, 309--54. El Colegio de México.
\url{https://doi.org/10.2307/j.ctv3dnrw3.15}.

\leavevmode\vadjust pre{\hypertarget{ref-stawarz2022}{}}%
Stawarz, Nico, Matthias Rosenbaum-Feldbrügge, Nikola Sander, Harun
Sulak, and Vanessa Knobloch. 2022. {``The Impact of the COVID{-}19
Pandemic on Internal Migration in Germany: A Descriptive Analysis.''}
\emph{Population, Space and Place} 28 (6).
\url{https://doi.org/10.1002/psp.2566}.

\leavevmode\vadjust pre{\hypertarget{ref-storper2004}{}}%
Storper, M., and A. J. Venables. 2004. {``Buzz: Face-to-Face Contact and
the Urban Economy.''} \emph{Journal of Economic Geography} 4 (4):
351--70. \url{https://doi.org/10.1093/jnlecg/lbh027}.

\leavevmode\vadjust pre{\hypertarget{ref-tatem2017}{}}%
Tatem, Andrew J. 2017. {``WorldPop, Open Data for Spatial Demography.''}
\emph{Scientific Data} 4 (1).
\url{https://doi.org/10.1038/sdata.2017.4}.

\leavevmode\vadjust pre{\hypertarget{ref-tuxf8nnessen2021}{}}%
Tønnessen, Marianne. 2021. {``Movers from the City in the First Year of
Covid.''} \emph{Nordic Journal of Urban Studies} 1 (2): 131--47.
\url{https://doi.org/10.18261/issn.2703-8866-2021-02-03}.

\leavevmode\vadjust pre{\hypertarget{ref-escap2019handbook}{}}%
UNSD. 2019. {``Handbook on the Use of Mobile Phone Data for Official
Statistics.''}

\leavevmode\vadjust pre{\hypertarget{ref-vogiazides2022}{}}%
Vogiazides, Louisa, and Juta Kawalerowicz. 2022. {``Internal Migration
in the Time of Covid: Who Moves Out of the Inner City of Stockholm and
Where Do They Go?''} \emph{Population, Space and Place} 29 (4).
\url{https://doi.org/10.1002/psp.2641}.

\leavevmode\vadjust pre{\hypertarget{ref-wang2022}{}}%
Wang, Yikang, Chen Zhong, Qili Gao, and Carmen Cabrera-Arnau. 2022.
{``Understanding Internal Migration in the UK Before and During the
COVID-19 Pandemic Using Twitter Data.''} \emph{Urban Informatics} 1 (1).
\url{https://doi.org/10.1007/s44212-022-00018-w}.

\end{CSLReferences}

\end{document}